\documentclass[pdflatex,referee,sn-nature]{sn-jnl}%
\usepackage{mathtools}
\usepackage{upgreek}
\usepackage{hhline}
\usepackage{tocbasic}
\DeclareNewTOC{stoc}
\newcommand\entrywithprefix[2]{#1~#2\hfill-\hfill}
\DeclareTOCStyleEntry[numwidth=6em, entrynumberformat=\entrywithprefix{\figurename}]{tocline}{figure}
\DeclareTOCStyleEntry[numwidth=6em, entrynumberformat=\entrywithprefix{\tablename}]{tocline}{table}

\usepackage{graphicx}
\usepackage{multirow}
\usepackage{amsmath,amssymb,amsfonts}
\usepackage{amsthm}
\usepackage{mathrsfs}
\usepackage[title]{appendix}
\usepackage{xcolor}
\usepackage{textcomp}
\usepackage{manyfoot}
\usepackage{booktabs}
\usepackage{algorithm}
\usepackage{algorithmicx}
\usepackage{algpseudocode}
\usepackage{listings}
\usepackage{tikz}
\usepackage{placeins}

\newcommand{\panelgraphic}[4][north west]{%
  \begin{tikzpicture}
    \node[anchor=south west, inner sep=0] (img) at (0,0) {\includegraphics[width=#2]{#3}};
    \begin{scope}[x={(img.south east)}, y={(img.north west)}]
      \foreach \px/\py/\pl in {#4} {
        \node[overlay, anchor=#1, inner sep=0, font=\sffamily\bfseries\small] at (\px,\py) {(\pl)};
      }
    \end{scope}
  \end{tikzpicture}%
}
\begin{document}

\title[Uncertainty in AI and Dynamical Climate Ensembles]{A Comparison of Uncertainty in AI and Dynamical Climate Ensembles}

\author*[1]{\fnm{Jacob B.} \sur{Landsberg}}\email{jlandsbe@bu.edu}
\equalcont{These authors contributed equally to this work.}

\author[1]{\fnm{Devin M.} \sur{McAfee}}\email{dmcafee@bu.edu}
\equalcont{These authors contributed equally to this work.}

\author[1]{\fnm{M.A.} \sur{Fernandez}}\email{martmart@bu.edu}
\equalcont{These authors contributed equally to this work.}

\author[1,2]{\fnm{Elizabeth A.} \sur{Barnes}}\email{eabarnes@bu.edu}

\affil[1]{\orgdiv{Faculty of Computing and Data Sciences}, \orgname{Boston University}, \orgaddress{\street{665 Commonwealth Ave}, \city{Boston}, \postcode{02215}, \state{MA}, \country{USA}}}
\affil[2]{\orgdiv{Department of Earth and Environment}, \orgname{Boston University}, \orgaddress{\street{685 Commonwealth Ave}, \city{Boston}, \postcode{02215}, \state{MA}, \country{USA}}}

\abstract{Climate model ensembles and their uncertainty estimates are essential for evaluating risk and designing climate policy. Yet, in the assessment of newly developed AI climate models, evaluating uncertainty is often secondary to diagnosing mean-state climate biases and trends. In this work, we conduct the first systematic comparison of uncertainty representation in ensembles of state-of-the-art AI and dynamical atmospheric climate models. Examining 2-meter temperature, 500 hPa geopotential height, and precipitable water, we find that the AI ensemble exhibits reduced internal variability and model uncertainty across variables relative to the dynamical ensemble. We show that this dampened variability causes the AI ensemble to produce more limited climate extremes and yields underdispersion relative to ERA5. Under a uniform warming scenario, the AI ensemble shows increased model uncertainty. Dynamical models grapple with their own challenges, exhibiting stronger miscalibration (overdispersion) relative to ERA5 and larger climate biases than the AI models. As AI and dynamical climate models are mutually integrated into operational climate prediction, careful consideration of their differences in uncertainty representations is crucial for grounded assessment of climate risk.}

\maketitle

\section{Introduction}\label{sec:intro}

Trustworthy climate uncertainty quantification is critical for assessing climate risks to human health, energy systems, agriculture, and infrastructure. For monthly-to-decadal climate predictions, uncertainty primarily stems from internal variability and model uncertainty \citep{hawkins2009}. Internal variability describes natural variations arising out of the chaotic nature of climate. Although it contains predictable components \citep{Mariotti2020}, internal variability dominates near-term climate uncertainty. Long-term climate uncertainty stems primarily from uncertainty about plausible climate change scenarios, as well as model uncertainty about the forced response in a given scenario \citep{hawkins2009}. Internal variability also plays a role for regional impacts at long timescales, especially at high latitudes \citep{deser2012uncertainty}. Internal variability, which may dampen or amplify the forced response, is crucial for assessing near-term climate risks, while model uncertainty, from computational limits and gaps in physical understanding, guides long-term climate planning. 
 
Ensembles of dynamical climate models, underpinned by governing physical equations, have long been the gold standard for modeling climate uncertainty \citep{IPCC_WG1,cmip6}. Climate ensembles consist of multiple independently perturbed simulations with identical forcing. In single-model climate ensembles, intramodel differences between simulations represent internal variability, while the ensemble mean approximates the forced response \citep{hawkins2009, Frankcombe2018}.

Multimodel climate ensembles comprise multiple single-model ensembles, so that model uncertainty can be evaluated through intermodel simulation differences \citep{lehner2020partitioning, IPCC_WG1}. Over recent decades, dynamical ensembles have formed the backbone of climate science, isolating mechanisms of climate variability and quantifying human influence on the climate system \citep{shaw2025other}. However, dynamical climate ensembles are restricted by their vast computational cost, which continues to grow with increasing model resolution and complexity \citep{Bordoni}. This cost limits the frequency at which dynamical ensembles can be produced and can prohibit the generation of extremely large ensembles required for sufficient representation of climate extremes \citep{Mahesh}. 

Recently, a new class of AI models, often referred to as AI emulators, has emerged as a potential tool for climate prediction at a fraction of the computational cost of dynamical models. AI emulators have achieved skill competitive with state-of-the-art dynamical weather models \citep{Lam, pangu, fourcastnet} and have recently expanded to climate prediction \citep{ace2,neuralgcm,arches1,arches2,dlesym,cbottle,md1p5}. AI climate emulators, which implement a diverse variety of model architectures, are typically designed to emulate climate reanalysis datasets, although efforts to develop models which emulate observational data are emerging \citep[e.g.][]{aardvark}.

While early results are promising \citep[e.g.][]{neuralgcm,ace2}, systematic analyses of the capabilities of AI emulators for climate uncertainty quantification are lacking. Weather emulators trained with deterministic losses are often underdispersive \citep[e.g.][]{Lam, Selz}, but it remains to be shown whether this underdispersion impacts the internal variability of climate emulators, and whether probabilistic losses remedy this behavior. Further, with the broadening horizon of climate emulation, multimodel AI ensembles are becoming feasible, but their uncertainty has yet to be fully evaluated. 

The first phase of the AI Model Intercomparison Project (AIMIP) was recently completed to address the need for comprehensive evaluation of AI climate emulators \citep{aimip_paper}. Following a consistent protocol, the AIMIP collaboration compiled the largest set of climate emulators for evaluation to date. Prior work \citep{aimip_paper,pallotta2026} has focused primarily on biases in forced variability within AIMIP. Here, we instead provide the first systematic comparison of uncertainty representation in AI models. Specifically, we compare representations of internal variability and model uncertainty in AIMIP with a multimodel dynamical ensemble comprised of a subset of CMIP6 AMIP (atmospheric model intercomparison project, \cite{cmip6}) simulations. We demonstrate how biases in these representations impact simulated historical variability. Moreover, we illustrate how uncertainty representation changes in AIMIP under an idealized climate change scenario -- an important test ahead of its potential integration into research and operational pipelines.

\section{Data \& Methods}\label{sec:data_methods}

\subsection{Data}\label{sec:data}

AIMIP \citep{aimip_paper} simulations are consistent with the AMIP protocol established for CMIP7 (coupled model intercomparison project phase 7, \cite{cmip7}). External forcing is provided through prescribed sea surface temperatures (SST) and sea ice concentrations (SIC) from monthly ERA5 \citep{era5} data that has been preprocessed for AIMIP\footnote{https://zenodo.org/records/17065758}. AIMIP models do not include CO$_2$ forcing due to concerns about overfitting \cite{aimip_paper}, so model generalizability to future climate is assessed through uniform +2 K SST and +4 K SST perturbation runs. 

AIMIP includes eight models: ArchesWeather and ArchesWeatherGen \citep{arches1,arches2}, ACE2.1-ERA5 \citep{ace2}, DLESyM \citep{dlesym}, NeuralGCM and NeuralGCM-HRD \citep{neuralgcm}, cBottle \citep{cbottle}, and MD-1p5 \citep{md1p5}. We exclude NeuralGCM as NeuralGCM-HRD represents the same model, but downscaled from 2.8-degree to the 1-degree resolution grid shared by the other contributing models (or equivalent for HEALPix grid models). Unless otherwise stated, we omit the fully-diffusion models (MD-1p5 and cBottle) from the main analysis. The fully-diffusion models exhibit undesirable artifacts such as deflated internal variability (Table \ref{tab:primary_var_table}) and, for MD-1p5, inflated temporal variance relative to the non-diffusion models. Further, there are five non-diffusion AIMIP models, which matches the size of the dynamical ensemble.

Each AI modeling center produced a five-member ensemble of monthly-mean outputs over a training period spanning 1979-2014 (36 years) and testing period spanning 2015-2024 (10 years). We do, however, also include annual results, computed using a rolling 12-month smoothing window in order to analyze variability at different time scales (e.g., Figure \ref{fig:smoothing}). Ensembles are generated by perturbing initial conditions, although there is no restriction on the perturbation method used. Further AIMIP data details, access instructions, and example processing code are available\footnote{https://github.com/ai2cm/AIMIP}.

\begin{table}[h!]
\caption{AIMIP models used in this study. Many of these models have been updated from their published design to accommodate the AIMIP protocol, e.g., the removal of CO2 as input, see \cite{aimip_paper}.}\label{table:aimodels}
\begin{tabular}{p{0.25\linewidth} | p{0.66\linewidth}}
Model Name & Notes\\
\midrule
ArchesWeather \citep{arches1} & Transformer architecture, autoregressive, RMSE loss. \\
\midrule
ArchesWeatherGen \citep{arches2} & Flow-matching model predicts residuals of the ArchesWeather deterministic output. \\
\midrule
ACE2.1-ERA5 \citep{ace2} & Spherical Fourier Neural Operator (SFNO) architecture, autoregressive, MSE loss. \\
\midrule
DLESyM \citep{dlesym} & U-net convolutional neural network architecture, autoregressive, RMSE loss, HEALPix grid \cite{healpix}. \\
\midrule
NeuralGCM-HRD \citep{neuralgcm} & Hybrid model (dynamical core with encoder-decoder architecture), autoregressive, CRPS loss. \\
\midrule
cBottle \citep{cbottle} & Diffusion architecture, not autoregressive, HEALPix grid. Ensembles generated using predictions from combinations of training checkpoints. \\
\midrule
MD1p5 \citep{md1p5} & Conditional variational autoencoder and latent diffusion architecture, autoregressive (monthly), MSE, KL regularization, and diffusion loss.\\
\botrule
\end{tabular}
\end{table}

We compare the AI models with a set of atmosphere-only dynamical models run under the AMIP protocol from CMIP6 \citep{cmip6}. We subselect models that contain at least five members for each of 2-meter temperature (TAS), geopotential height at 500 hPa (Z500), and specific humidity for the period 1979-2014. While AIMIP models' test set extends to 2024, the AMIP experiment only consistently extend to 2014, which limits the time range of our comparison. The dynamical models we use in this work are: CESM2 (10 members), GISS-E2-1-G (20 members), IPSL-CM6A-LR (21 members), MIROC6 (10 members), and NorCPM1 (10 members). These dynamical models are regridded to match the one-degree grid of the AIMIP models using conservative regridding. Henceforth, we refer to this dynamical ensemble as DENS. 

In this work, we evaluate TAS, Z500, and lower-free-tropospheric precipitable water (PRW). PRW is calculated by integrating specific humidity between 500 hPa and 850 hPa, where high altitude regions that extend into this column have been omitted. We choose to examine this subsection of the full atmospheric column, as (1) not all models have data at other pressure levels and (2) moisture in the lower free troposphere is physically important for the onset of deep convection, especially in the tropics \citep{GlobalTropicalPrecipitationRelationshipstoFreeTroposphericWaterVaporUsingRadioOccultations}. DLESyM does not include specific humidity, so it is absent from PRW analysis. For TAS, we mask out the oceans, as TAS is tightly coupled to the prescribed sea surface temperatures with which the models are forced. We focus our comparison on 1979-2014, as this is the common period between AIMIP and DENS. We also include analysis of the AIMIP test period (2015-2024), as well as the uniform +2 Kelvin SST simulations, to examine out-of-sample behavior in AIMIP. 

\subsection{Methods}\label{sec:methods}

For each model within AIMIP and DENS, the model’s time-mean is removed at every grid point to filter the influence of mean-state biases. We compute variances at each monthly time step, both across members within each model (atmospheric internal variability) and across model means (model uncertainty). This formulation is standard for climate ensembles, and assumes that internal variability and model uncertainty are independent and can be represented as additive sources of variance \citep[e.g.,][]{lehner2020partitioning}. The remaining variance of the multimodel ensembles is explained by the time-variance of the atmospheric forced response (atmospheric forced variability). The decomposition of each framework’s variance into contributions from atmospheric internal and forced variability, while taking into account model uncertainty, allows us to explain discrepancies in the variability between the two frameworks through differences in each component of variance. Importantly, although this formulation of variance decomposition is common \citep{lehner2020partitioning,hawkins2009}, it overestimates model uncertainty and atmospheric forced variability with small sample sizes \citep{QuantifyingSourcesofUncertaintyinProjectionsofFutureClimate}. Model means and multi-model means both retain sampling noise, which thereby inflates differences between models (model uncertainty) and time variance (atmospheric forced variability). A random-effects model that corrects for this yields smaller estimates of both variances, but does not change the comparison between AIMIP and DENS and is discussed in Supplemental Section \ref{sec:supp_biascorr}.  

Throughout the manuscript, we refer to atmospheric internal variability more generally as internal variability, and atmospheric forced variability as forced variability. These notions should not be confused with the internal and forced components of coupled climate ensembles. Specifically, in coupled ensembles, the forced response measures the climate system's response to radiative forcing. Here, we use forced response to refer to the atmosphere's response to SST forcing. Variability of this forced response across monthly time steps is mostly due to seasonality; while on interannual to decadal time scales, the atmospheric forced response is influenced by internal variability of the SSTs, e.g., ENSO, as well as local SST trends due to climate change.

Confidence of the internal variability and model uncertainty estimates are evaluated by bootstrapping on time, using leave-one-model-out, and for DENS models, subsampling on member. Within each temporal bootstrap, internal variability and model uncertainty are computed by randomly sampling, with replacement, from the 1979-2014 period. When computing model uncertainty, within each temporal bootstrap we randomly subsample (20 times, without replacement) 5 members from each DENS model (which have up to 21 members) to match the ensemble size of the AIMIP models. This is necessary to ensure that the model uncertainty of DENS is not artificially reduced due to lower uncertainty about the forced response \citep[e.g.][]{Frankcombe2018}. We average over the time bootstrap dimension (and member for DENS model uncertainty), with the number of iterations chosen based on convergence of the bootstrap process (see Supplemental Figure \ref{fig:interconverge}). As we are primarily interested in comparing the full DENS and AIMIP frameworks, not individual models, we also use leave-one-model-out to provide uncertainty in our framework-averaged results (e.g., Table \ref{tab:primary_var_table}). %

Beyond comparing the frameworks' variances, we also evaluate their extremes and marginal calibrations. To analyze extremes, we calculate spatial quantiles by pooling all times and members of each model -- these are the quantiles defined by the member-mean empirical cumulative distribution function (CDF). Bootstrapping on quantiles was performed in the same way as the model uncertainty bootstrapping, but without leave-one-model-out, as it made little difference to the distribution of quantiles. To assess the marginal calibration of each framework, we compute the probability integral transform (PIT) of ERA5 reanalysis using the multimodel distributions of AIMIP and DENS. 

For each framework at gridpoint $i$, we pool times, members (for DENS, the first five members), and models to form a multimodel empirical CDF $F_i$. PIT values are calculated as $F_{i}(\text{ERA5}_{i, t})$ for each ERA5 sample across time at gridpoint $i$. We note that PIT values are traditionally calculated using time-conditional forecast distributions $F_{i, t}(\text{ERA5}_{i, t})$ as a test of probabilistic calibration for weather models \citep[e.g.,][]{Gneiting2007}. Here, we are interested in the marginal calibration of climate models, so we scrutinize the climatological CDF $F_{i}$. For a marginally calibrated ensemble, the PIT values will approximately follow a standard uniform distribution. Dispersion can be assessed by comparing the variance of the PIT values to the variance of the standard uniform, which is $1/12$. An underdispersed ensemble will have PIT variance $> 1/12$, while an overdispersed ensemble will have PIT variance $< 1/12$ \citep{gneitingdispersion}. We quantify dispersion in Figure \ref{fig:dispersion} as $\log(12 \times \text{Var}(\text{PIT}))$, the log-ratio of the PIT variance to the standard uniform variance. This dispersion metric is negative for overdispersion and positive for underdispersion.

Global calibration across gridpoints is examined using PIT histograms created by partitioning the PIT values into decile bins. A latitude-weighted average of each gridpoint's PIT histogram is computed to produce the global PIT histograms reported in Figure \ref{fig:pit}. To measure deviations from marginal calibration, we report the D-metric (see \citep{Nipen} and Section \ref{sec:dmetric_bootstrap}).

\section{Results}\label{sec:results}

\begin{figure}[h!]
  \centering
  \panelgraphic[north east]{0.95\textwidth}{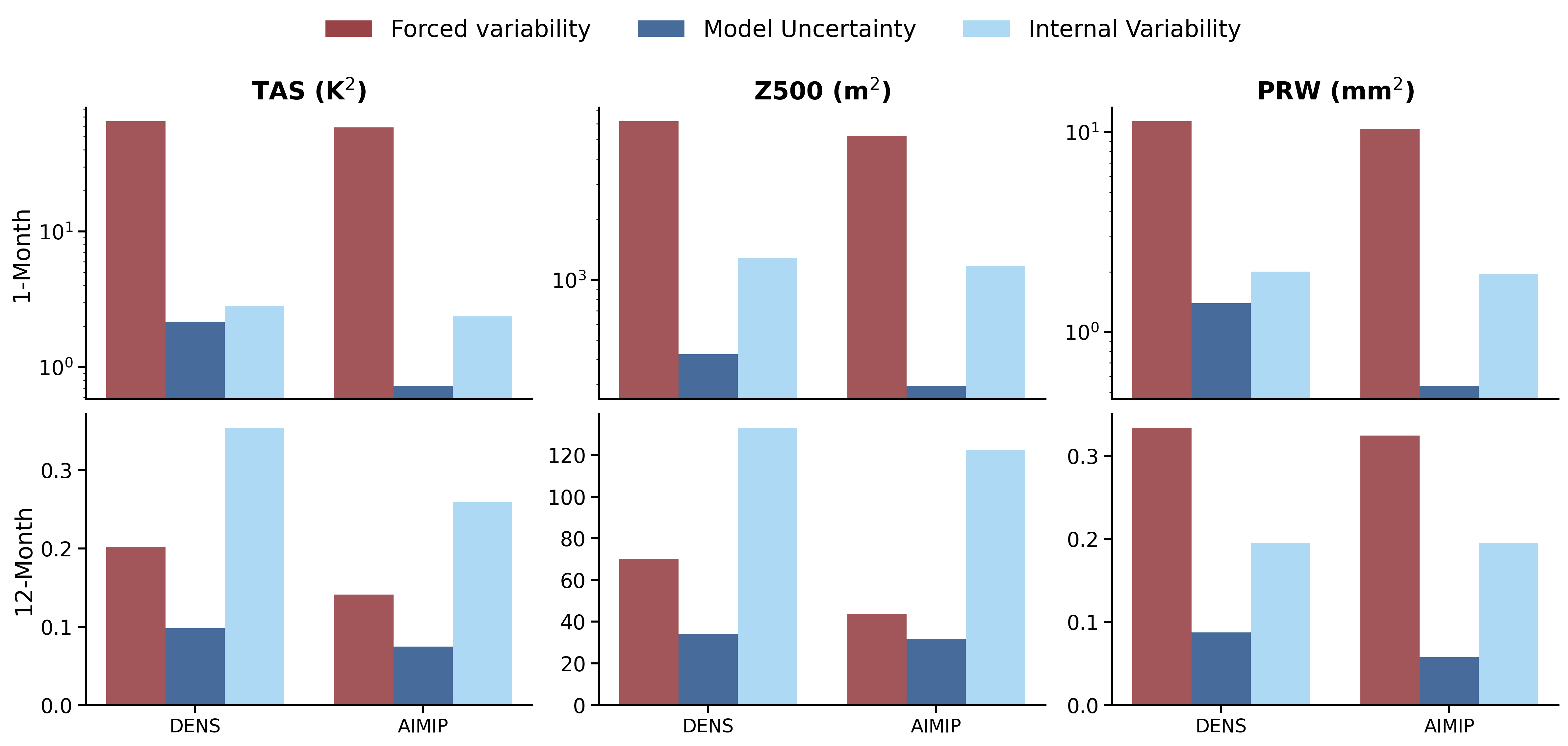}{-0.004/0.862/a, -0.004/0.443/b}
  \caption{Internal variability, model uncertainty and time-variance of the multimodel-mean (forced variability) for AIMIP and DENS over the training period. Both (a) monthly and (b) annual (12-month smoothed) results are shown; the monthly y-axis is on a log scale, so that differences in internal variability and model uncertainty are visible.} \label{fig:smoothing}
  \end{figure}

Mean monthly internal variability and model uncertainty across frameworks for TAS, PRW, and Z500 are reported in Table \ref{tab:primary_var_table}, along with confidence intervals defined by the range over bootstrap and leave-one-model-out iterations. We find that DENS simulates stronger internal variability and exhibits greater model uncertainty than non-diffusion AIMIP models across variables (Table \ref{tab:primary_var_table}), although this difference in internal variability does not reach the level of significance for PRW. Further, while not always significant, DENS shows similarly larger internal variability and model uncertainty on the annual timescale. This is also observed in the annually smoothed simulations (Figure \ref{fig:smoothing}). Moreover, this disparity grows for the two diffusion models, which show significantly weaker internal variability than all of the non-diffusion models across variables (Table \ref{tab:primary_var_table}, see also Table \ref{tab:intra_by_model}).

\begin{table*}[h!]
  \caption{Internal variability and model uncertainty for DENS and AIMIP models. As there are only two AIMIP diffusion models, we do not report a model uncertainty for the diffusion framework. Bolded values indicate confidence intervals that do not overlap with the AIMIP non-diffusion models over the training period.}
  \label{tab:primary_var_table}
  \includegraphics[width=0.95\textwidth]{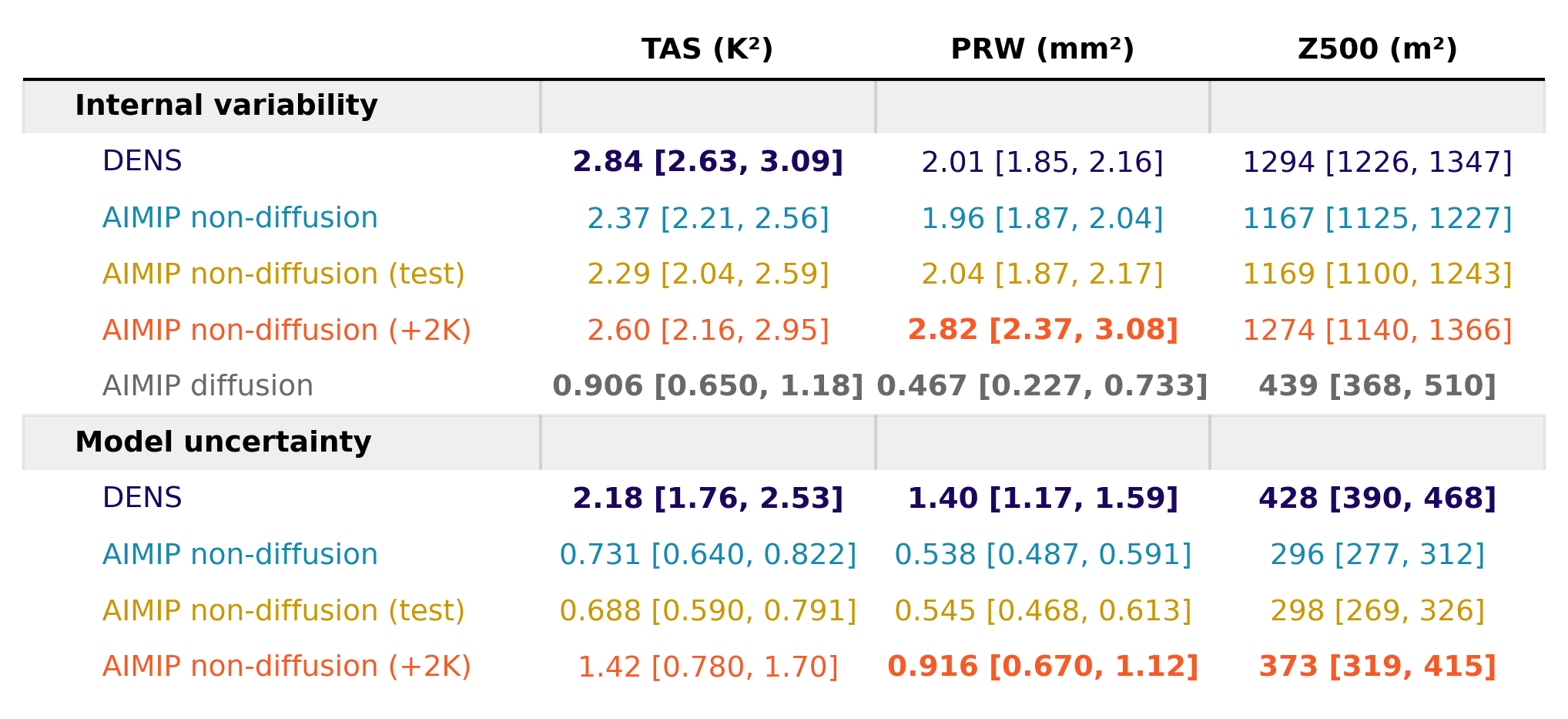}
\end{table*}

Over the test period, AIMIP's internal variability and model uncertainty remain roughly unchanged. On the other hand, model uncertainty noticeably increases under the idealized +2 K scenario. This is perhaps expected, as the AIMIP models produce a range of unphysical responses to the out-of-sample forcing, such as muted TAS sensitivity globally and the inability to capture the land-sea warming contrast. For instance, ACE2.1-ERA5 and cBottle exhibit significant out-of-sample artifacts, which simulate a near-surface cooling response across much of the global landmass \citep{aimip_paper}. The internal variability of AIMIP also increases, but only significantly so for PRW. The modest increases in internal variability for TAS and Z500 shown in Table \ref{tab:primary_var_table} are due to the unphysical response of a single model, ACE2.1-ERA5, whose members diverge across variables (Supplemental Table \ref{tab:intra_by_model}). However, the increase in internal variability of PRW is significant and is observed in all of the non-diffusion AIMIP models, not only ACE2.1-ERA5 (see Supplemental Table \ref{tab:intra_by_model}). ACE2.1-ERA5 is similarly responsible for much of the increase in PRW model uncertainty in the +2K scenario.

In Figure \ref{fig:smoothing}, we show forced variability, as well as time-mean internal variability and model uncertainty for AIMIP and DENS at monthly and annual time scales.  At the monthly time scale, across variables, variance is dominated by forced variability (Figure \ref{fig:smoothing}a), largely due to the effect of seasonality, as the contribution from forced variability decreases considerably under annual smoothing (Figure \ref{fig:smoothing}b). For AIMIP, model uncertainty contributes substantially less variance than internal variability at the monthly time scale. For DENS, however, model uncertainty is on par with the magnitude of internal variability for monthly TAS and PRW. Presumably, this difference in model uncertainty is because the AIMIP models were all trained to emulate ERA5's seasonal cycle. Therefore, AIMIP has comparatively less model uncertainty about its seasonal cycle, resulting in reduced monthly model uncertainty relative to DENS. 

At the annual time scale, seasonality is removed, leading to a reduction in total variance (Figure \ref{fig:smoothing}b). Internal variability is the largest source of the remaining variance for TAS and Z500 in both AIMIP and DENS. However, forced variability remains the primary contributor to PRW variance annually. This is likely caused by the fact that the global hydrological cycle is dominated by the moist tropics, particularly the tropical Pacific \citep[e.g.,][]{Trenberth2007}; and tropical Pacific variability is modulated by ENSO, which has peak strength at the interannual time scale \citep{Timmermann2018}. Broadly, DENS has stronger forced variability than AIMIP at both time scales in Figure \ref{fig:smoothing}. However, unlike TAS and Z500, on annual timescales DENS and AIMIP have nearly identical PRW forced variability. We speculate that this is also related to ENSO. While the AIMIP models generally have weaker sensitivity to the prescribed SSTs globally, they have a stronger PRW response primarily confined to the tropical Pacific (see Supplemental Figures \ref{fig:supp_prw_extremes} and \ref{fig:supp_prw_annual_extremes}). While model uncertainty is still comparatively larger for DENS at the annual time scale, the relative difference between the frameworks' uncertainties is smaller compared to the monthly simulations. In other words, while AIMIP has less uncertainty about the seasonal cycle, AIMIP's uncertainty regarding longer term forced variability is more comparable to the dynamical ensemble.

\begin{figure}[ht!]
\centering
\panelgraphic{0.70\textwidth}{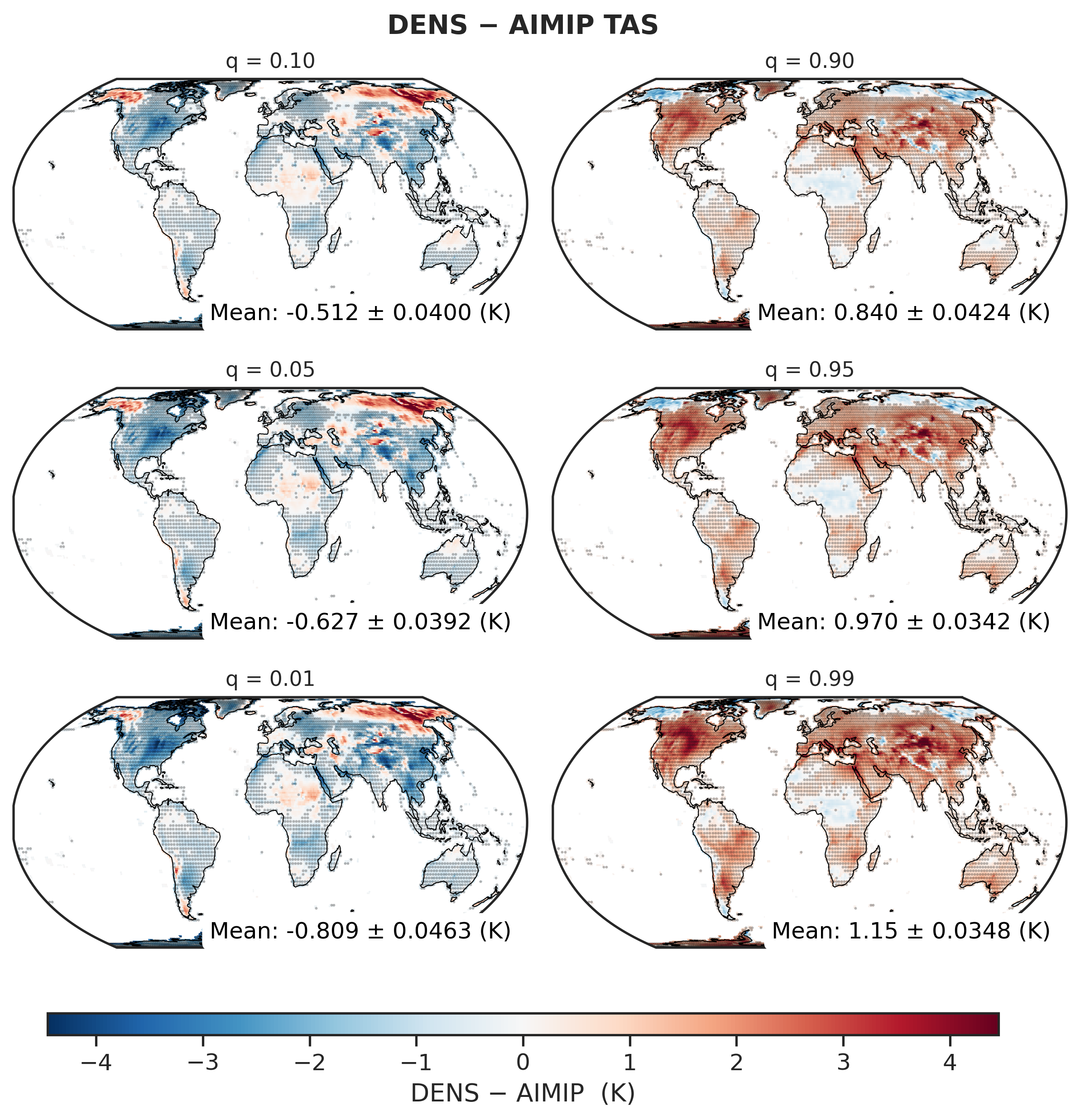}{0/1/a}
\panelgraphic{0.65\textwidth}{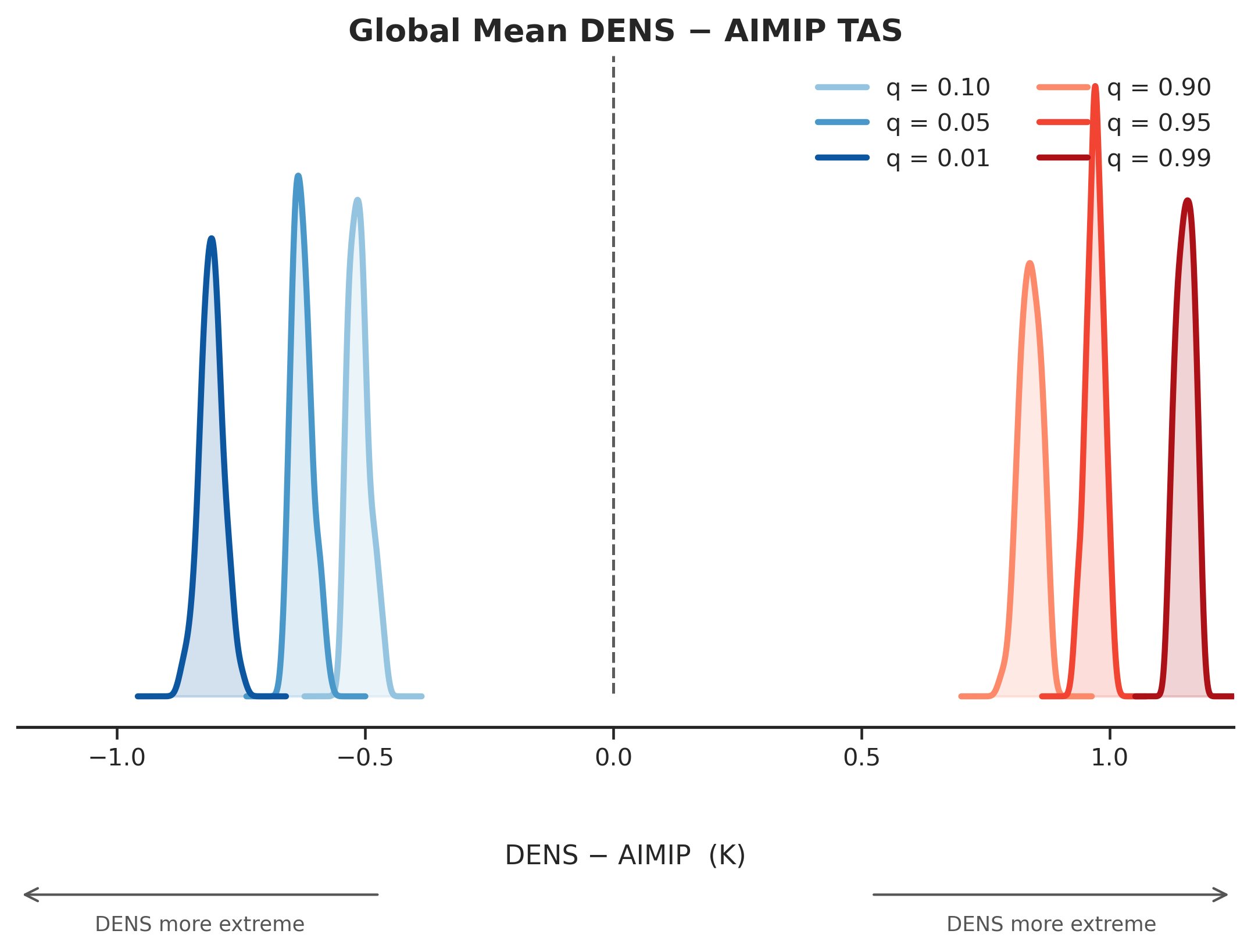}{0/1/b}
\caption{(a) Global-mean differences in 1st, 5th, 10th, 90th, 95th, and 99th quantiles between DENS and AIMIP for TAS.
Stippling indicates DENS values are more extreme than those in AIMIP for all bootstrap iterations.
(b) Global mean of the differences from panel (a).
Distributions and uncertainty ($2\sigma$) are computed from bootstrapped distributions.}\label{fig:extremes}
\end{figure}

We now illustrate how monthly differences in components of variability between the two frameworks manifest in extremes representations. In Figure \ref{fig:extremes} we show the difference between the 1st, 5th, 10th, 90th, 95th, and 99th percentiles (averaged across models) between DENS and AIMIP for monthly TAS (PRW and Z500 shown in Supplemental Figures \ref{fig:supp_prw_extremes} and \ref{fig:supp_z500_extremes}). We find that DENS produces greater extremes than AIMIP and that the discrepancy between frameworks is largest for the most extreme quantiles. For example, for TAS, the 99th percentile of TAS in DENS is more than 1 K hotter than that of AIMIP, while the 95th percentile is just under 1 K hotter (Figure \ref{fig:extremes}b).

For monthly timescales, this difference is primarily influenced by the weaker seasonality of AIMIP, as monthly forced variability is significantly stronger than internal variability for both frameworks (Figure \ref{fig:smoothing}). A Shapley value analysis attributes, for example, 86\% of the monthly TAS $q=0.99$ discrepancy to differences in forced variability (see Supplemental section \ref{supp:sec_shapley}). As shown in Figure \ref{fig:extremes}a, near-surface temperature extremes are stronger in DENS than AIMIP over the majority of the global landmass. This TAS difference is most prominent in the northern extratropics, and comparatively smaller in the southern hemisphere, consistent with the northern hemisphere's stronger seasonal temperature variability \citep{Donohoe2013}. Similarly, Z500 extremes are weaker over most regions in AIMIP. However, the largest Z500 differences occur in the extratropics, where seasonal variability is strongest (Supplemental Figure \ref{fig:supp_z500_extremes}). 

The discrepancies in monthly PRW extremes are more spatially non-uniform. Over most of the extratropics, AIMIP is less dispersive and yields weaker global-mean quantiles. However, AIMIP produces stronger PRW extremes mostly over the central and eastern tropical Pacific (Supplemental Figure \ref{fig:supp_prw_extremes}), where forced variability is dominated by ENSO. For instance, over the NINO3.4 region \citep{Barnston1997}, $q = 0.99$ is 2.58 mm higher in AIMIP -- the latitude-weighted global-mean difference is $-0.37$ mm -- with 2.13 mm attributable to differences in forced variability (Supplemental Figure \ref{fig:shapley_global_monthly}). 

We observe a similar discrepancy for the annual time scale, with AIMIP producing increasingly weaker extremes than DENS as quantiles become more extreme (see Supplemental Figures \ref{fig:supp_tas_annual_extremes}, \ref{fig:supp_prw_annual_extremes}, and \ref{fig:supp_z500_annual_extremes}). Since annual variance is dampened relative to the monthly simulations, the annual quantile differences are comparatively smaller. Internal variability has a greater contribution to the differences in TAS extremes compared to the monthly time scale, where seasonality dominates. For example, the global-mean $q = 0.99$ is 0.27 K colder in AIMIP for annual TAS, with 53\% of the difference attributable to internal variability (see Supplemental Figure \ref{fig:shapley_global_annually}). However, for Z500 and PRW, differences in single-model forced variability between frameworks still play a dominant role (see Supplemental section \ref{supp:sec_shapley}). 

\begin{figure}[h!]
\centering
\panelgraphic{0.9\textwidth}{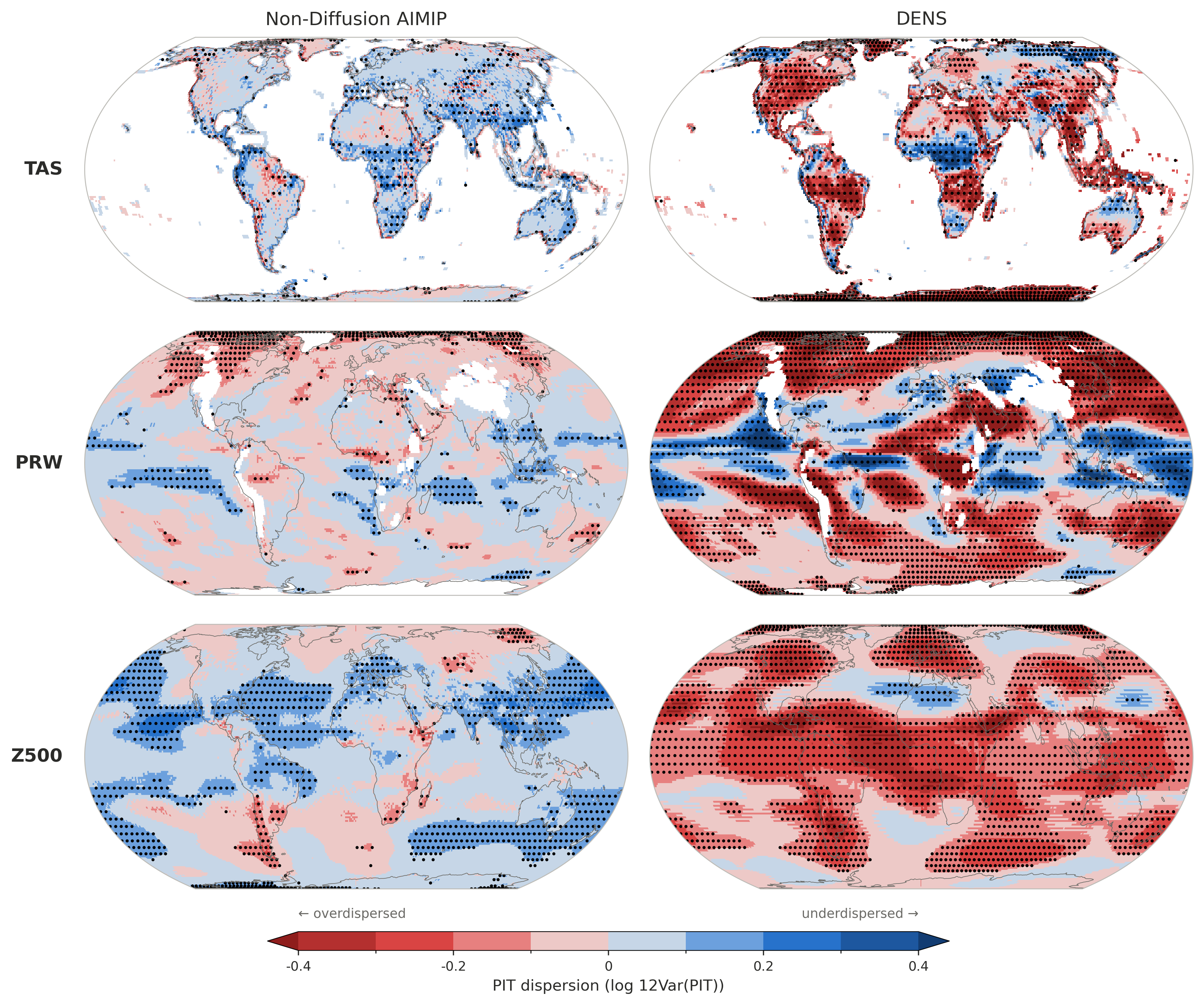}{0.07/0.965/a, 0.54/0.965/b, 0.07/0.672/c, 0.54/0.672/d, 0.07/0.379/e, 0.54/0.379/f}
\caption{PIT dispersion ratio for (a) AIMIP TAS, (b) DENS TAS, (c) AIMIP PRW, (d) DENS PRW, (e) AIMIP Z500, and (f) DENS Z500 over the training period.
Stippling denotes significance ($p \leq 0.05$) using a two-tailed Monte-Carlo test, where the null distribution is generated by calculating the variance of $N_{\text{eff}}$ samples from a standard uniform and repeating for 1000 iterations; $N_{\text{eff}}$ denotes the number of effective samples for estimating the variance of the PIT values at a given gridpoint, accounting for temporal autocorrelation \citep{Vehtari2021}.}\label{fig:dispersion}
\end{figure}

It is evident that the multimodel AIMIP ensemble is less dispersive than DENS. However, while informative, this comparison is limited without any relation to observational or reanalysis data. Thus, we assess the marginal calibration of both frameworks against ERA5 using PIT analysis. In Figure \ref{fig:dispersion}, we show spatial maps of PIT dispersion for each framework across variables at the monthly timescale. DENS is significantly overdispersed in each variable for most regions of the globe, with the exception of PRW over much of the tropical basins. Local calibration of AIMIP is superior to DENS, with far fewer regions reaching significance. However, AIMIP is underdispersed, particularly in TAS (Figure \ref{fig:dispersion}a) and Z500 (Figure \ref{fig:dispersion}e), while exhibiting overdispersion for polar PRW -- which is perhaps due to the minimal climatological PRW of the dry polar regions. While dispersion in AIMIP PRW is spatially nonuniform over the training period, AIMIP is more uniformly underdispersed in PRW over the testing period, while remaining underdispersed in TAS and Z500 (see Supplemental Figure \ref{fig:dispersion_test}).

Global PIT histograms are shown in Figure \ref{fig:pit}, with D-metrics provided in Supplemental Table \ref{tab:pitd_bootstrap}. Marginal calibration is signified by a uniform PIT histogram, with each bin containing an equal fraction of ERA5 samples. For an underdispersed ensemble, ERA5 samples will disproportionately occupy the extreme tails of the ensemble's marginal distribution, producing a U-shaped histogram; conversely, the PIT histogram of an overdispersed ensemble will form an inverted U.

The global monthly PIT histograms (Figure \ref{fig:pit}) paint the same picture as the spatial maps: DENS exhibits overdispersion globally (significant D-metric for Z500), while AIMIP showcases comparatively weaker underdispersion (significant D-metric for TAS). For Z500, the biases in DENS are relatively symmetric for the high and low quantiles, but the biases in TAS and PRW are skewed towards an overrepresentation of warm and dry extremes (Figure \ref{fig:pit}b). Over the testing period, the AIMIP non-diffusion models are underdispersed relative to ERA5 across the three variables. However, with a limited sample relative to the training period, the D-metrics do not achieve significance (see Supplemental section \ref{sec:dmetric_bootstrap}). 

At the annual time scale, AIMIP remains underdispersed in TAS and Z500, exhibiting underdispersion in TAS over the vast majority of global landmass (see Supplemental Figure \ref{fig:dispersion_12month}). As a result, global PIT histograms exhibit significant deviations from uniformity for all three variables at the annual time scale for AIMIP (Supplemental Figure \ref{fig:diff_pit_12_month}). Conversely, annual Z500 and PRW are overdispersed in DENS over much of the globe; however, the TAS dispersion signal in DENS is much more spatially heterogenous than at the monthly time scale (Supplemental Figure \ref{fig:dispersion_12month}) and does not reach significance globally (Supplemental Figure \ref{fig:diff_pit_12_month}). 

\begin{figure}[h!]
\centering
\panelgraphic[north east]{0.75\textwidth}{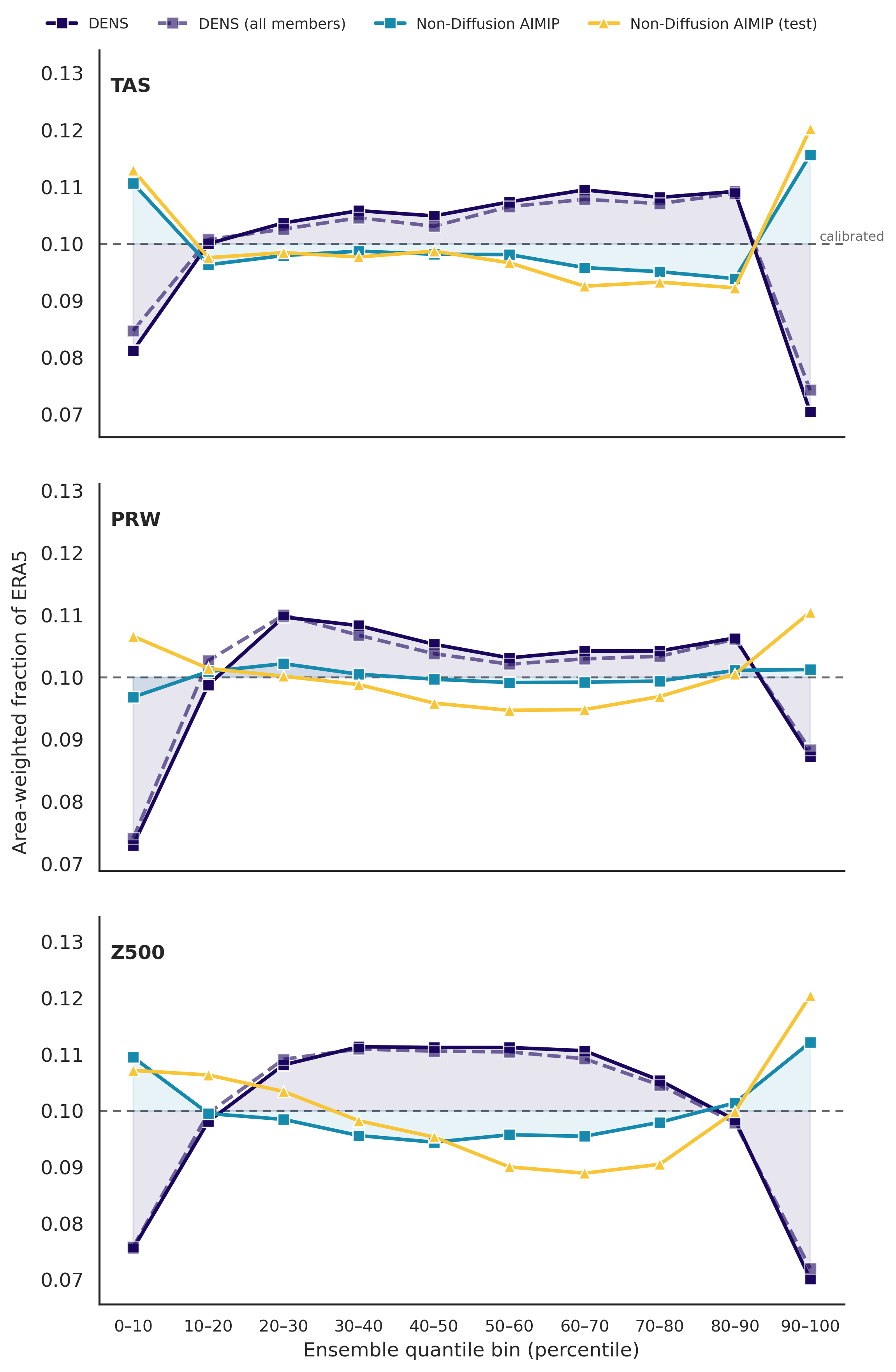}{-0.005/0.965/a, -0.005/0.648/b, -0.005/0.332/c}
\caption{Global probability integral transform (PIT) histograms for TAS, PRW, and Z500 for DENS and non-diffusion AIMIP models. A perfectly calibrated model would produce a uniform PIT histogram.}\label{fig:pit}
\end{figure}

\FloatBarrier %
\section{Discussion \& Conclusion}\label{sec:disc_conc}

In this study, we provide the first systematic comparison of uncertainty representations between AMIP-style multimodel ensembles of AI climate models (AIMIP) \citep{aimip_paper} and dynamical models. We find that the AI ensemble produces weaker forced variability, as well as reduced internal variability and model uncertainty relative to their dynamical peers at both monthly and annual time scales. As a result, climatological extremes in AIMIP are milder than the dynamical ensemble in TAS, Z500 and PRW, with the exception of PRW extremes over the tropical oceans. This suppressed variability implies that the AIMIP models produce a narrower range of climate states than their dynamical counterparts, while also exhibiting unphysical responses to out-of-sample idealized warming scenarios, potentially inhibiting their utility for climate risk assessment.

When compared to ERA5 on monthly time scales, AIMIP exhibits modest underdispersion in its climatological distributions of Z500 and TAS. In contrast, the dynamical ensemble shows overdispersion for all three variables which is stronger than the underdispersion signal of AIMIP. On annual time scales, AIMIP remains underdispersed across all three variables as compared to monthly, while the dynamical ensemble shows insignificant miscalibration for TAS, but significant overdispersion for PRW and Z500. The superior marginal calibration of AIMIP with respect to ERA5 is perhaps expected, as the AI models were designed to emulate ERA5 over the training period. Further, AIMIP exhibits similar or stronger underdispersion in the testing period compared to the training period. Thus, both frameworks suffer from miscalibrated uncertainty representation, but in different ways and to different degrees, making the choice between current AI emulators and dynamical models use-case dependent.

AI climate models offer clear advantages over physical models, achieving low climatological biases with realistic ENSO representations, at a fraction of the computational cost \cite{aimip_paper}. Conversely, AI models across different disciplines \cite{Ying_2019, medical_overfitting_ex, language_model_overfitting} can overfit to their training data. Previous work has found that AI climate emulators underpredict historic warming trends \cite{aimip_paper, singh2026evaluatingskillstabilityarchesweather, under_predicting_parts_of_dist} and struggle under future climate scenarios \cite{ace2}. In this work we show that, despite often exhibiting bettter calibration than a dynamical ensemble relative to ERA5, a multimodel ensemble of AI models (AIMIP) nonetheless underestimates variability and uncertainty.

Current AI models are predominantly optimized to improve predictive skill and reduce mean biases. However, their uncertainty quantification needs to be calibrated and trustworthy. As the community pushes towards the integration of AI climate models into research and operational pipelines, this work demonstrates the need to prioritize efforts, especially in the context of multimodel ensembles, on improving representations of internal variability and model uncertainty in AI models.

\backmatter

\section*{Data availability}
AIMIP model outputs and accompanying ERA5 data are available following the instructions provided at: \url{https://github.com/ai2cm/AIMIP/blob/main/evaluations/README.md}.
DENS model outputs are taken from the Pangeo CMIP6 Google Cloud Storage AMIP experiments and were accessed using Intake-ESM (\url{https://intake-esm.readthedocs.io/en/stable/index.html}).

\section*{Code availability}
The underlying code for all the analyses presented will be made available upon publication.

\bibliography{sn-bibliography}%

\section*{Acknowledgements}
This work was supported, in part, by the Global Water Security Center through DoD HQ00342420014, grant \#DE-SC0024212 from the U.S. Department of Energy, grant \#AGS-2210068 from the U.S. National Science Foundation, grant \#NA24OARX431C0022 from the National Oceanic and Atmospheric Administration (NOAA) and the Program for Coordinated Model Diagnosis and Intercomparison (PCMDI) Earth System Modeling project, which is supported by the Earth and Environmental Systems Modeling program in the U.S. Department of Energy Office of Science.

\section*{Author contributions}
J.B.L., D.M.M., and M.A.F. contributed to the conceptualization of the work, methodology, data acquisition and processing, analysis, and writing.
E.A.B. contributed to writing/editing and suggestions about analysis.

\section*{AI Disclosure}
Claude Code (Anthropic) was used to assist in generating the code for the experiments described in this project, as well as scripts for creating publication figures. The authors reviewed and verified the content of both the code repository and associated manuscript text and take responsibility for their accuracy and correctness.

\section*{Competing Interest}
The authors declare no competing interests.

%% ---------------- Supplementary Information ----------------
\clearpage
\setcounter{section}{0}\setcounter{figure}{0}\setcounter{table}{0}\setcounter{equation}{0}
\renewcommand\thesection{S\arabic{section}}
\renewcommand\thefigure{S\arabic{figure}}
\renewcommand\thetable{S\arabic{table}}
\renewcommand\theequation{S\arabic{equation}}
\renewcommand\theHsection{S\arabic{section}}
\renewcommand\theHfigure{S\arabic{figure}}
\renewcommand\theHtable{S\arabic{table}}
\renewcommand\theHequation{S\arabic{equation}}
\let\mainaddcontentsline\addcontentsline
\renewcommand{\addcontentsline}[3]{\mainaddcontentsline{stoc}{#2}{#3}}

\begin{center}
    {\LARGE \bf Supporting Material for: A Comparison of Uncertainty in AI and Dynamical Climate Ensembles}
\end{center}
\vspace{-0.5cm}

\begin{center}
    {\large Jacob B. Landsberg (\url{jlandsbe@bu.edu}), Devin M. McAfee, M.A. Fernandez,  and Elizabeth A. Barnes}
\end{center}

\listoftoc*{stoc}
\clearpage

\section{Model Biases}\label{sec:bias}

\begin{table*}[ht!]
    \centering
    \caption[Mean Bias for TAS, PRW, and Z500]{Mean bias of TAS, PRW, and Z500, relative to ERA5, for DENS and AIMIP.}
    \label{table:bias}
    \includegraphics[width=0.85\textwidth]{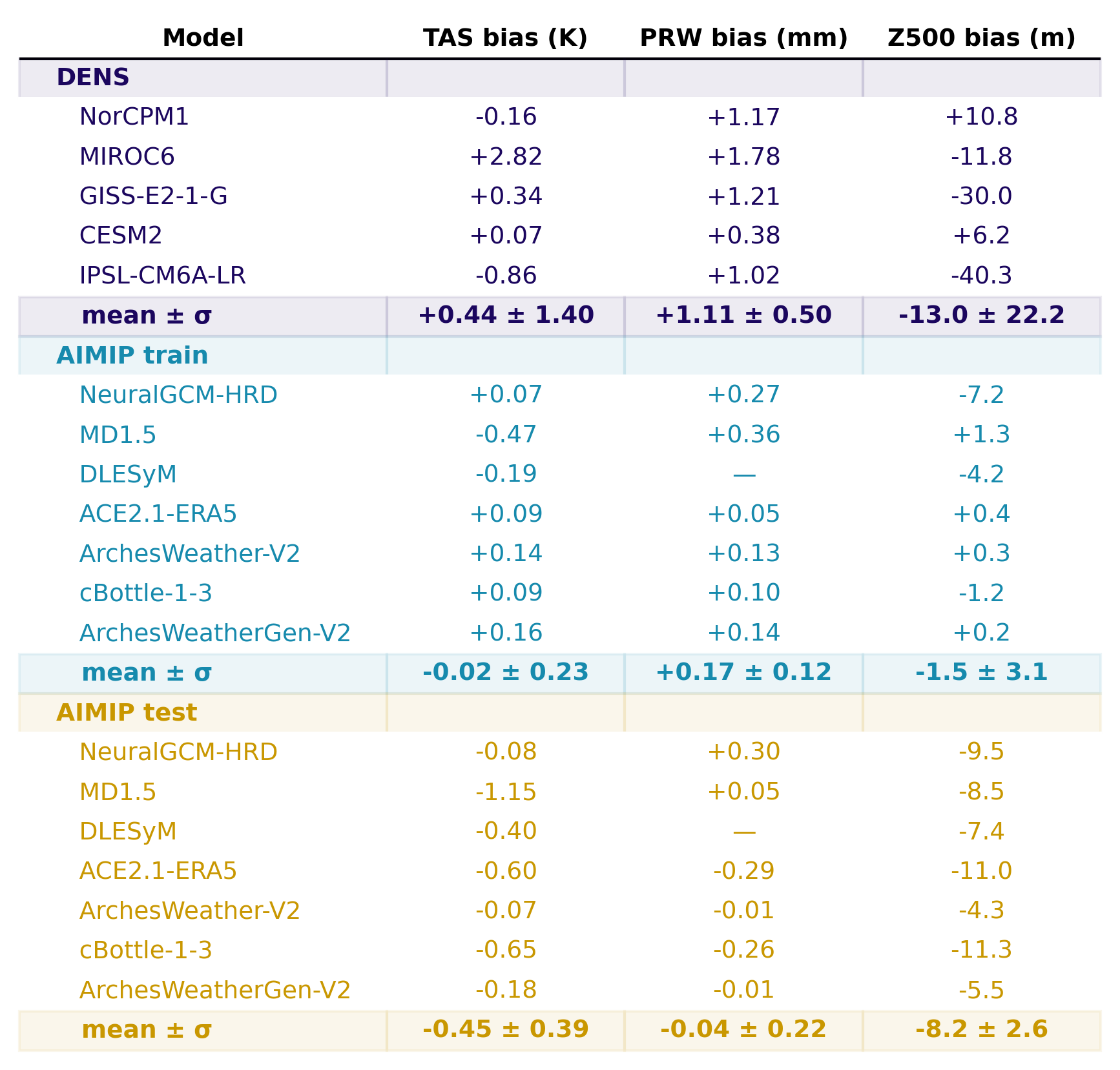}
\end{table*}
\clearpage

\section{Internal Variability and Model Uncertainty}\label{sec:intra_inter_maps}

\begin{table*}[ht!]
  \centering
  \caption[Internal Variability by AIMIP Model]{Monthly global-mean internal variability for each AIMIP model over the training period (1979--2014), test period (2015--2024), and +2\,K simulation, given as the mean over temporal bootstraps. Bolded values indicate where the bootstrap ranges (minimum to maximum) do not overlap between the model's training period and its testing period or +2 K simulation.}
  \label{tab:intra_by_model}
  \includegraphics[width=0.85\textwidth]{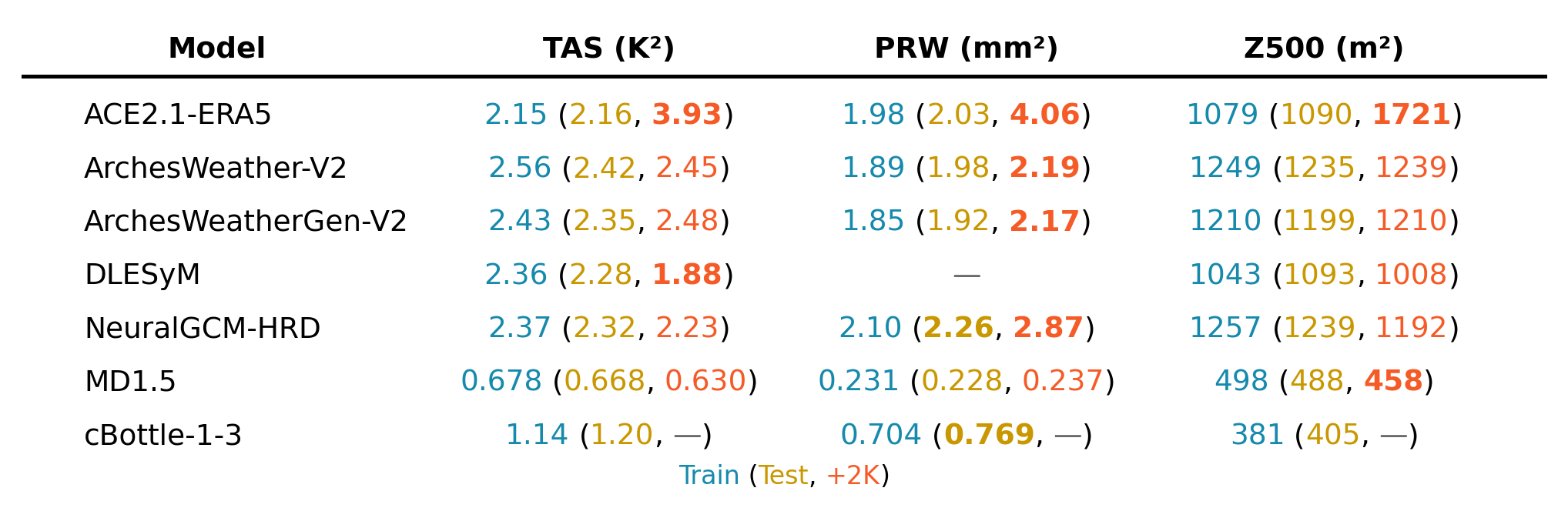}
\end{table*}

\begin{figure}[ht!]
  \centering
  \includegraphics[width=0.95\textwidth]{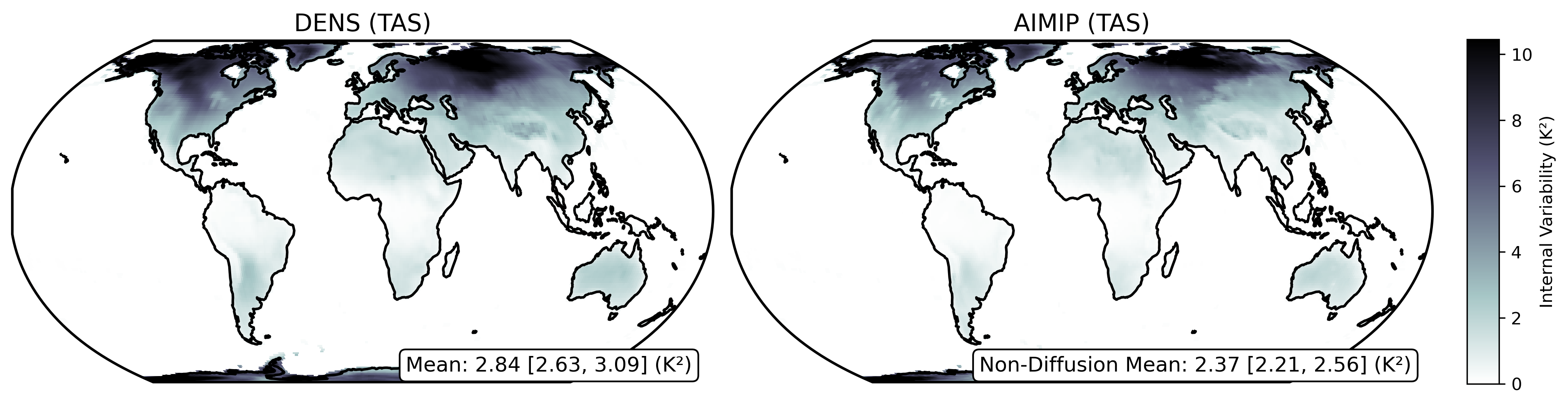}
  \includegraphics[width=0.95\textwidth]{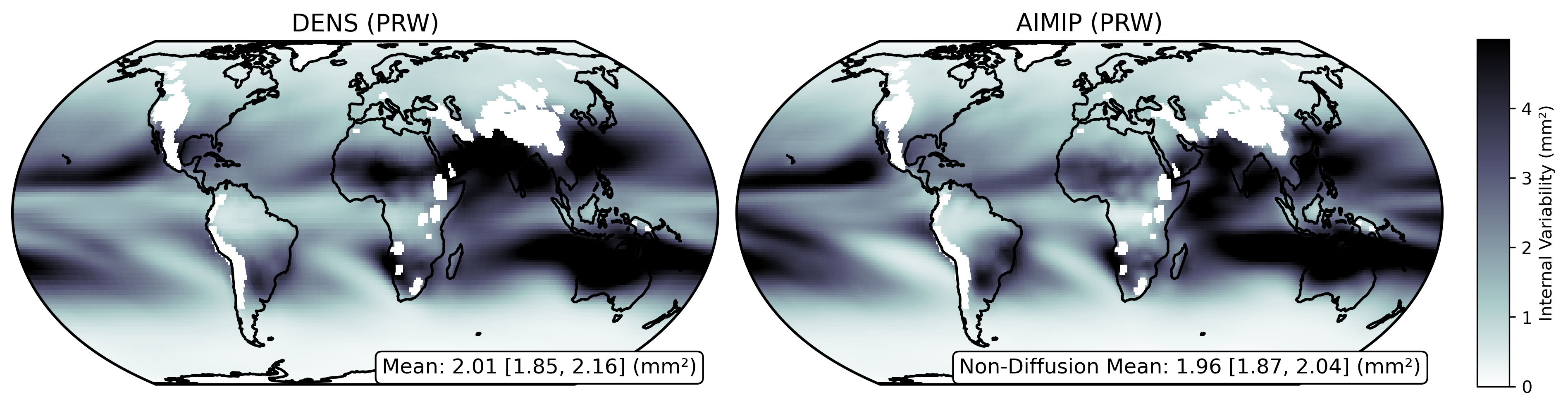}
  \includegraphics[width=0.95\textwidth]{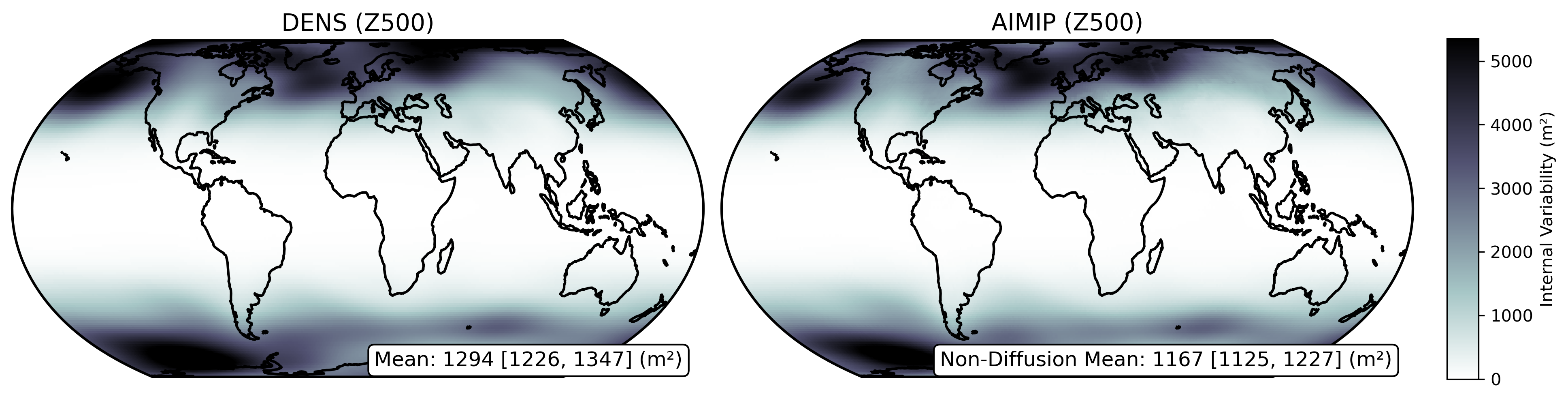}
  \caption[Internal Variability Variance Maps]{Internal variability of DENS and AIMIP models for the variables TAS, PRW, and Z500. For TAS, the oceans have been masked to avoid excess influence from the model forcing, which is primarily driven by prescribed sea surface temperatures. For all variables, the internal variability of DENS is larger than AIMIP.}\label{fig:intramodel}
\end{figure}

\begin{figure}[h!]
  \centering
  \includegraphics[width=0.95\textwidth]{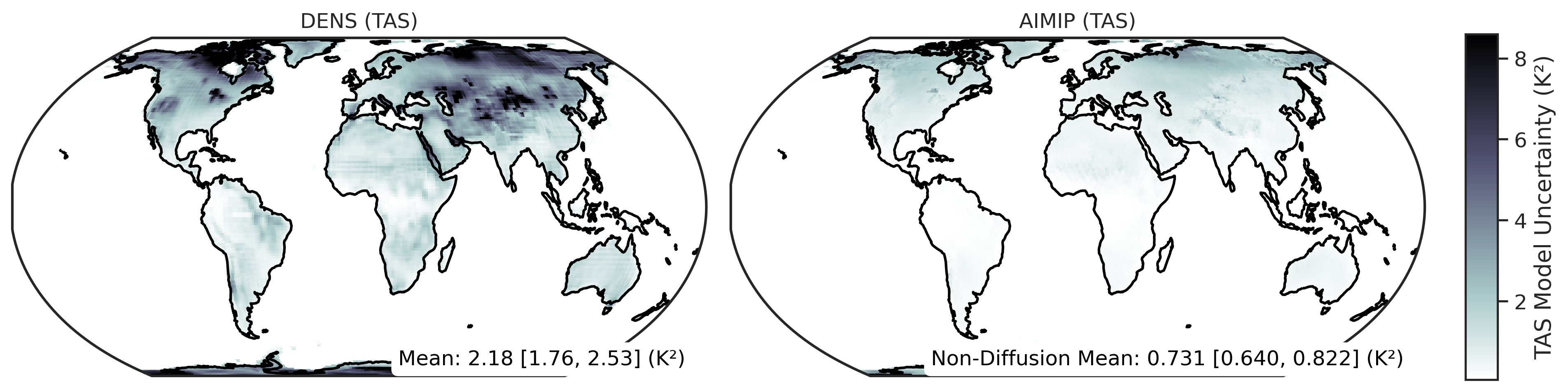}
  \includegraphics[width=0.95\textwidth]{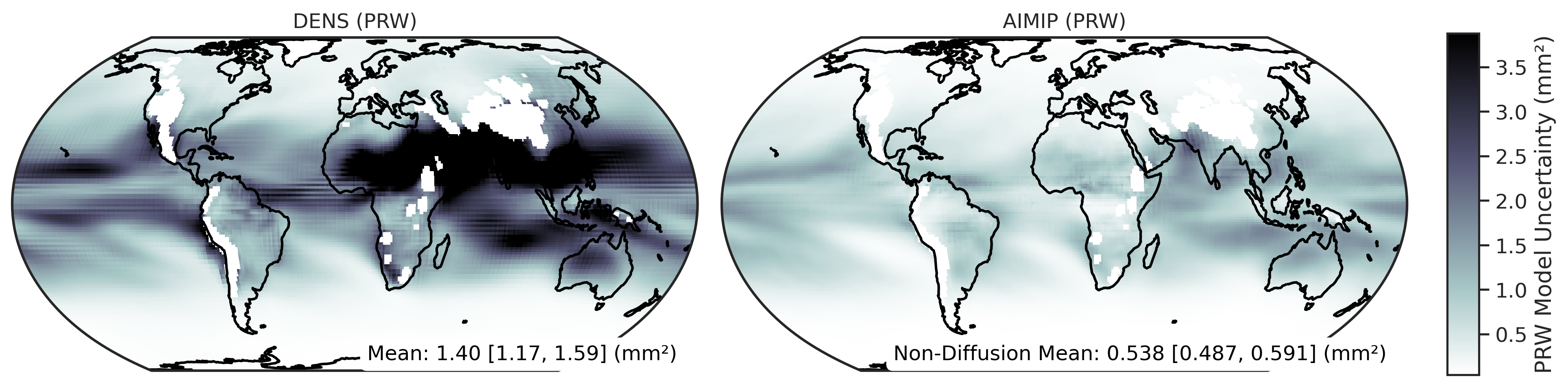}
  \includegraphics[width=0.95\textwidth]{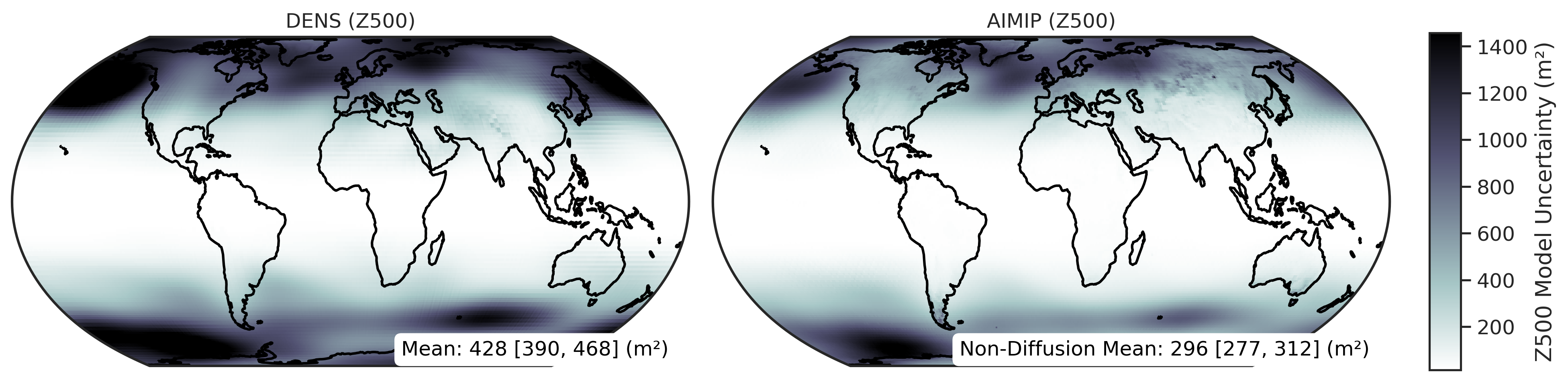}
  \caption[Model Uncertainty Maps]{Model uncertainty of DENS and AIMIP models for the variables TAS, PRW, and Z500. For TAS, the oceans have been masked to avoid excess influence from the model forcing, which is primarily driven by prescribed sea surface temperatures. For all variables, the model uncertainty of DENS is larger than AIMIP.}\label{fig:intermodel}
\end{figure}
\clearpage

\section{Bootstrap Convergence}\label{sec:bsconverge}

\begin{figure}[ht!]
    \centering
    \includegraphics[width=0.85\textwidth]{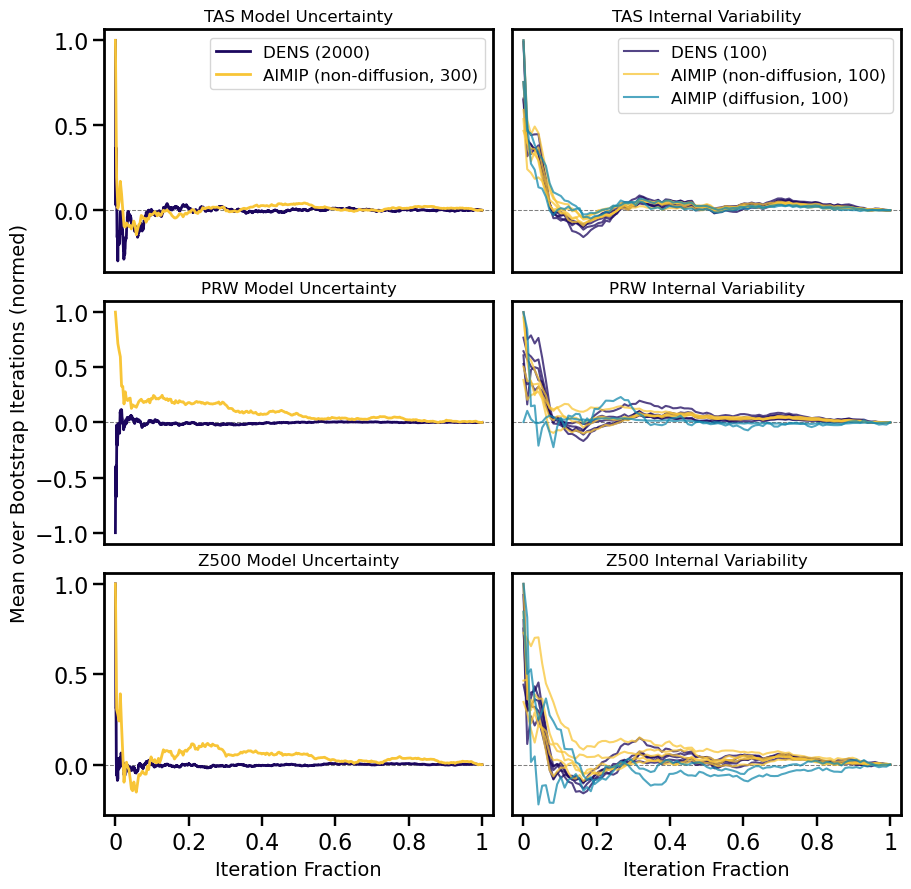}
    \caption[Intermodel and intramodel convergence]{Convergence of intermodel and intramodel variance bootstrapping.
    For each variable (TAS, PRW, Z500), the mean over all bootstrap iterations up to that iteration is shown (number of iterations shown in brackets).
    Values are averaged over time, latitude, longitude, and model within that framework.
    The left panels show the intermodel variance convergence of DENS models, which includes an extra subsampling step to account for the different number of members in each model, and the AIMIP non-diffusion models.
    The right panels show intramodel variance convergence, with different colors corresponding to different samples of models.}\label{fig:interconverge}
\end{figure}
\clearpage

\section{Extremes}\label{sec:supp_extremes}

\begin{figure}[ht!]
  \centering
  \includegraphics[width=0.70\textwidth]{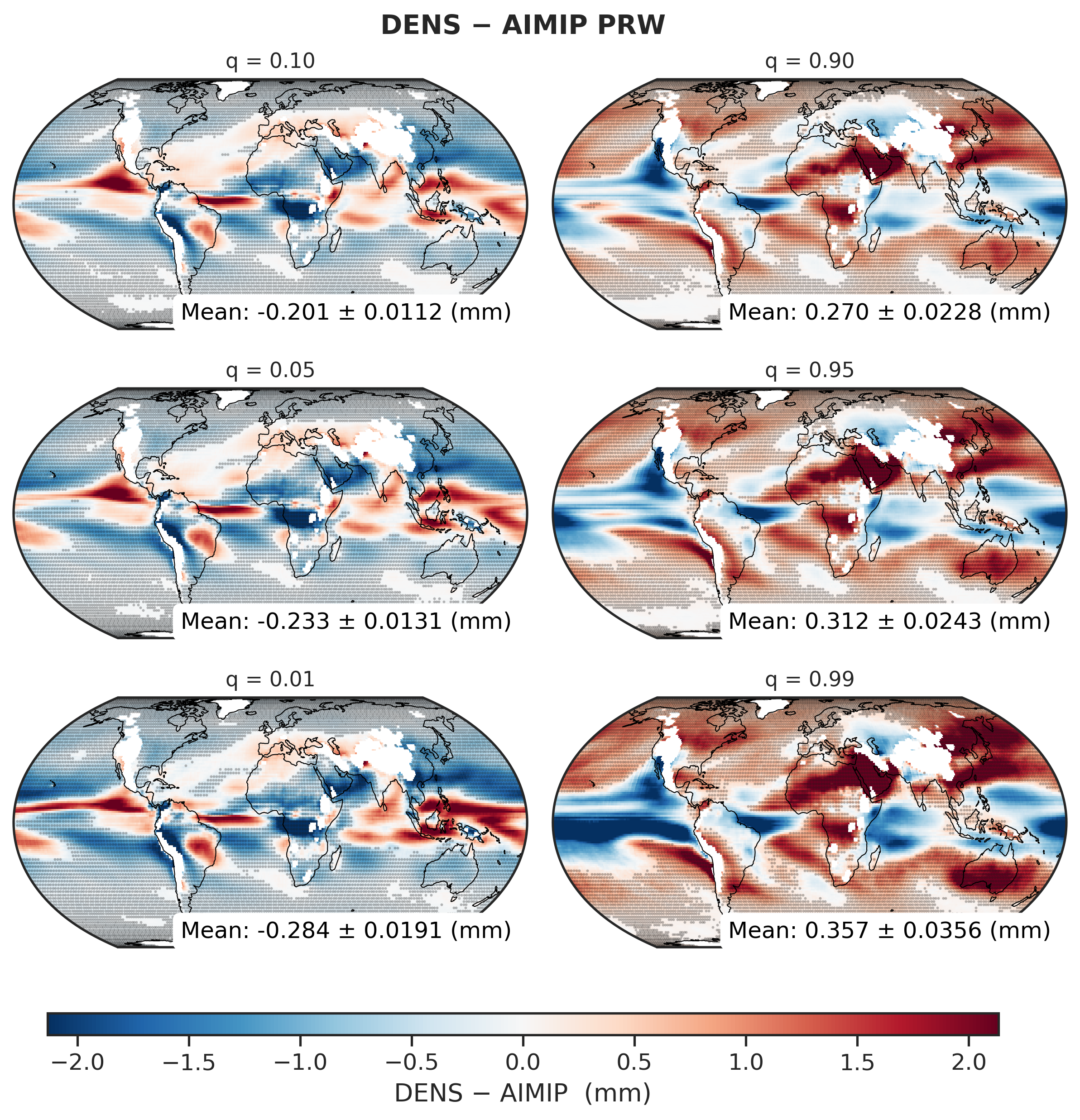}
  \includegraphics[width=0.65\textwidth]{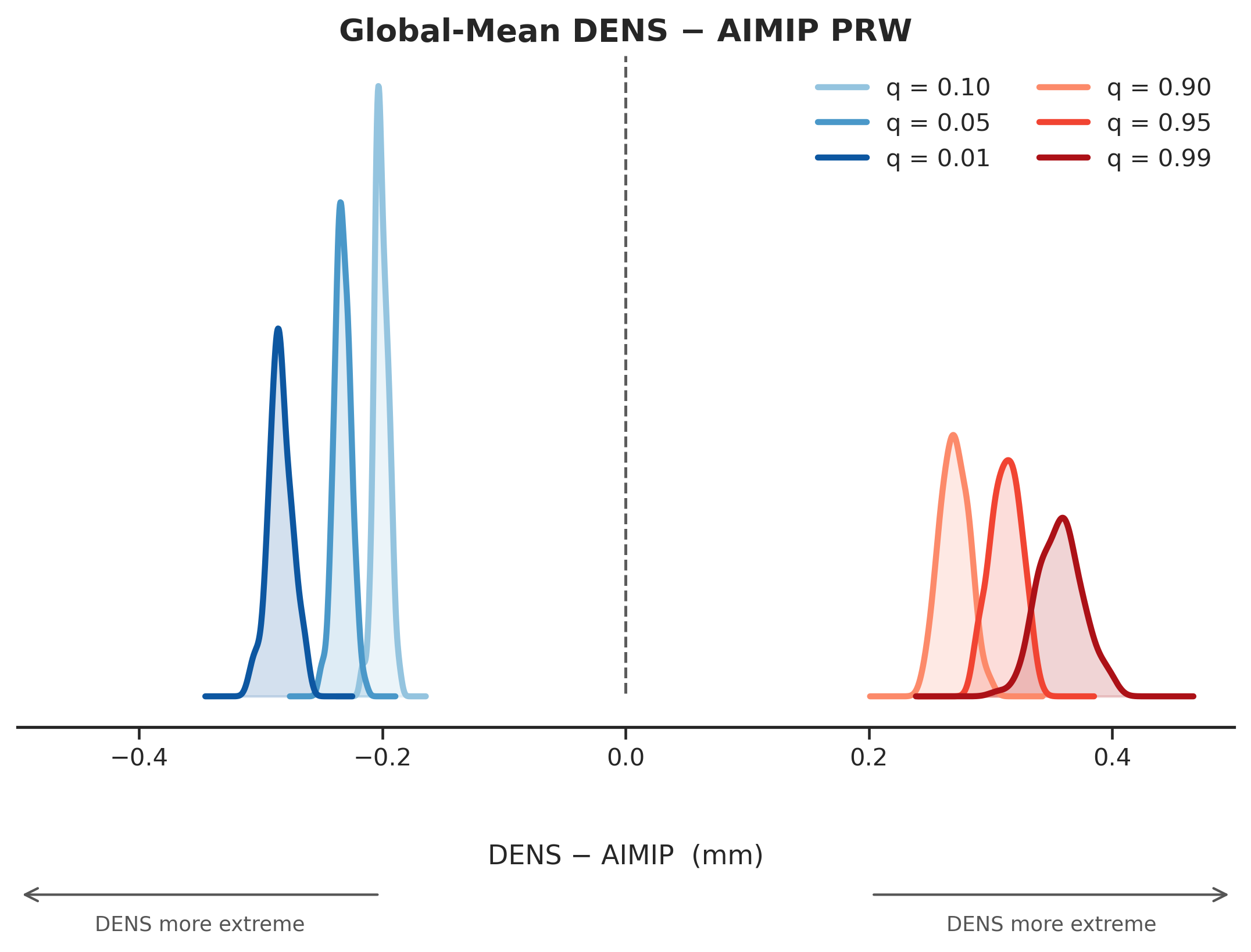}
  \caption[Map and distribution of monthly PRW extreme differences]{Same as Figure \ref{fig:extremes}, but for PRW.}
  \label{fig:supp_prw_extremes}
\end{figure}

\begin{figure}[ht!]
  \centering
  \includegraphics[width=0.70\textwidth]{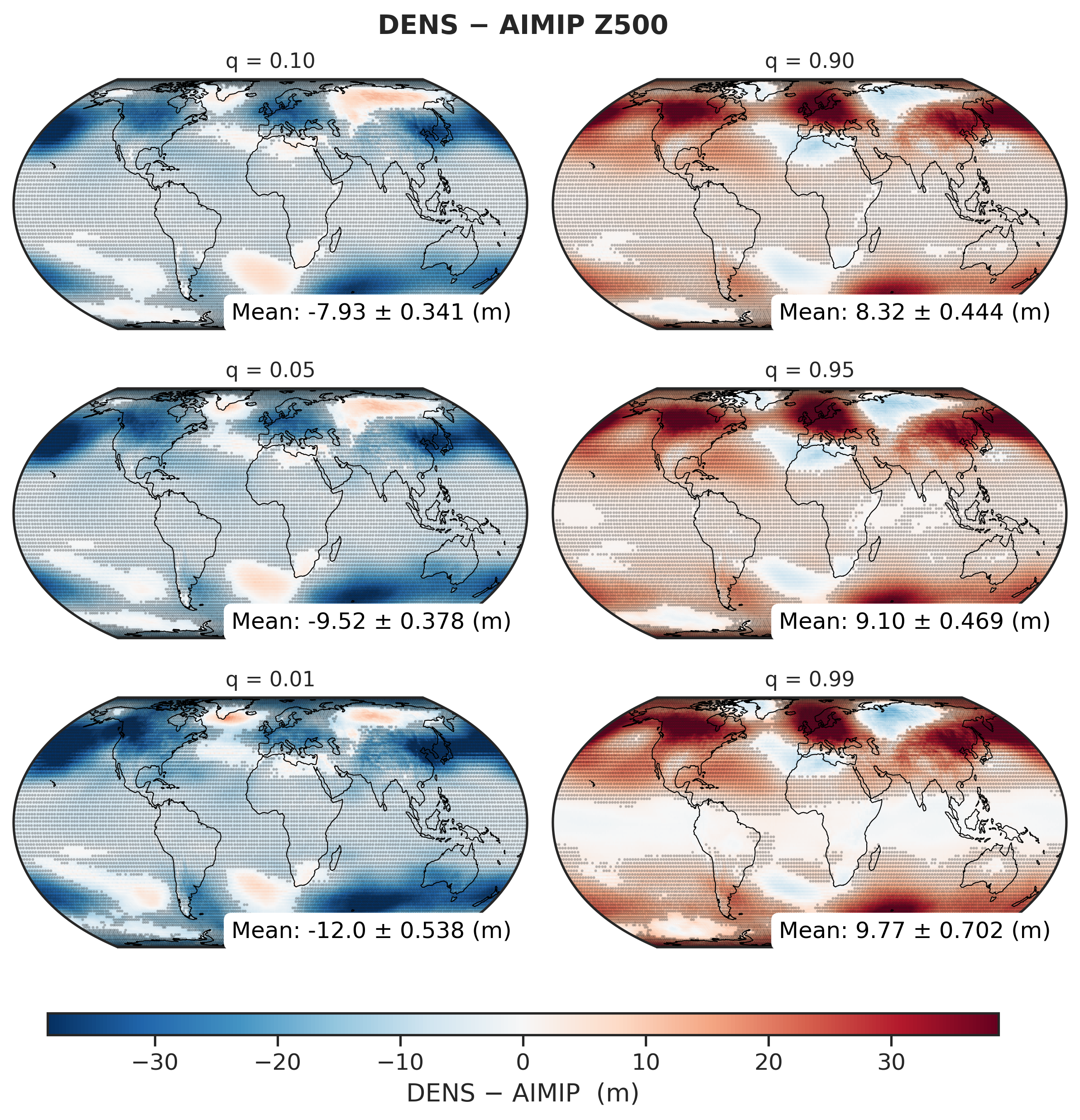}
  \includegraphics[width=0.65\textwidth]{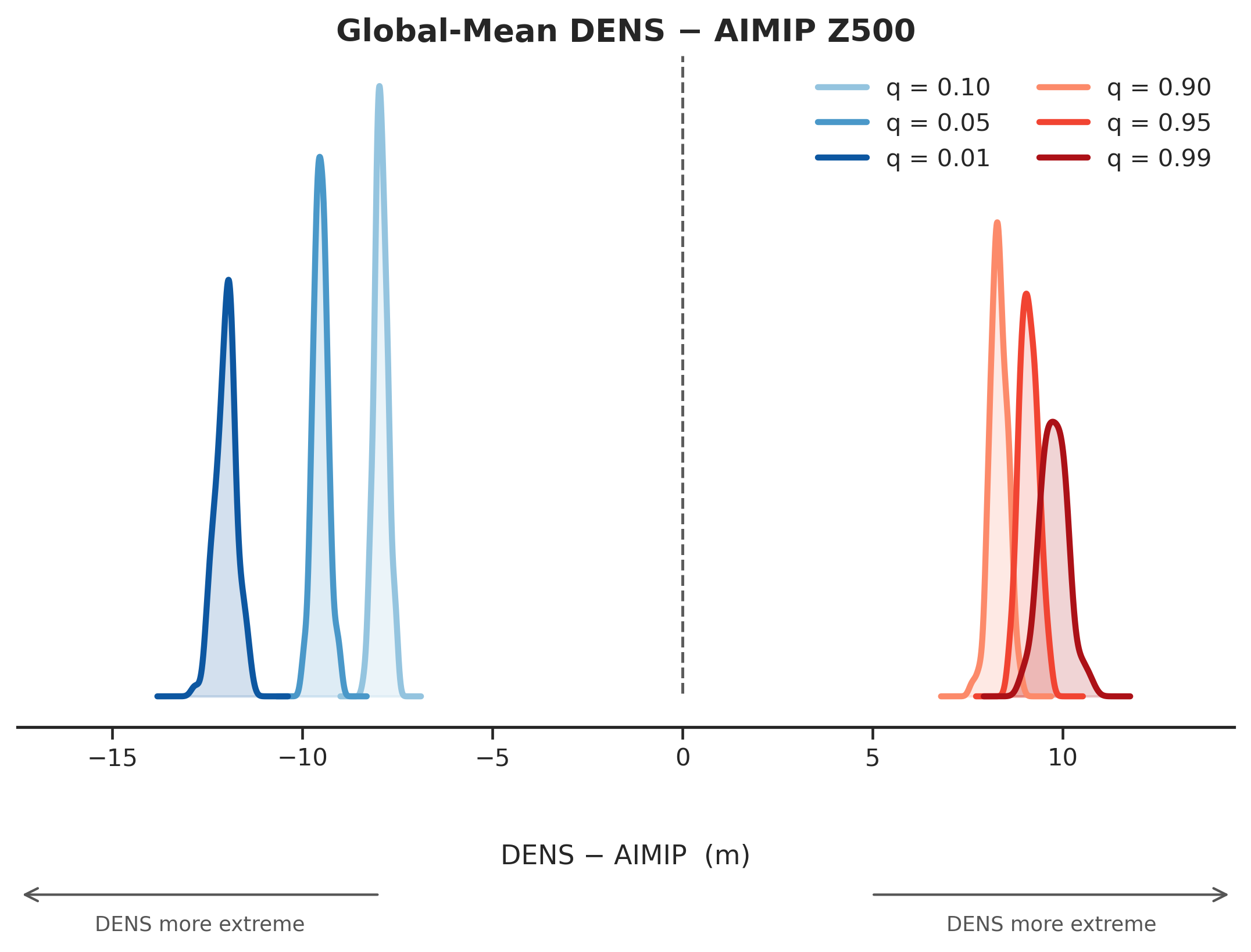}
  \caption[Map and distribution of monthly Z500 extreme differences]{Same as Figure \ref{fig:extremes}, but for Z500.}
  \label{fig:supp_z500_extremes}
\end{figure}

\begin{figure}[ht!]
  \centering
  \includegraphics[width=0.70\textwidth]{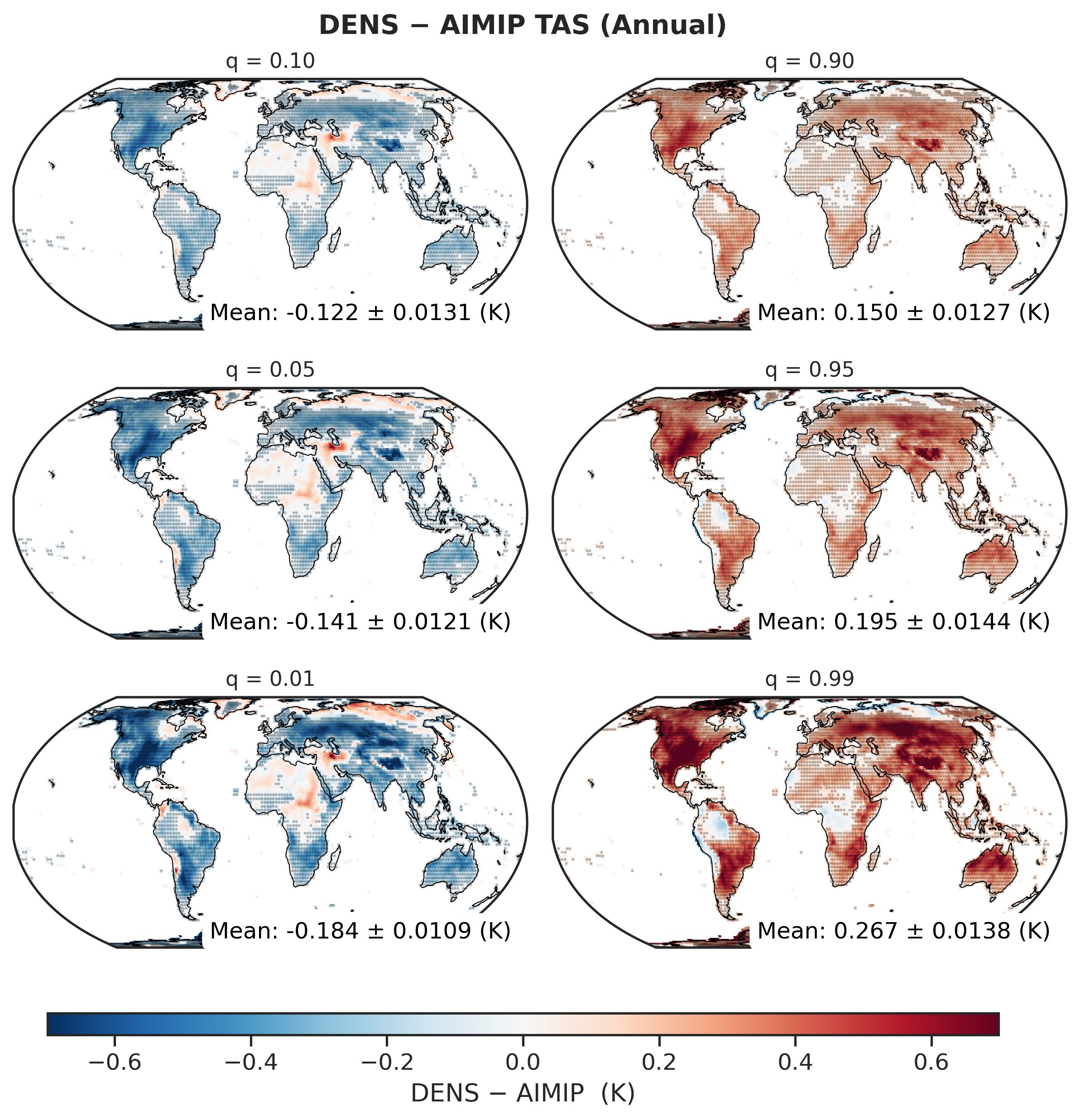}
  \includegraphics[width=0.65\textwidth]{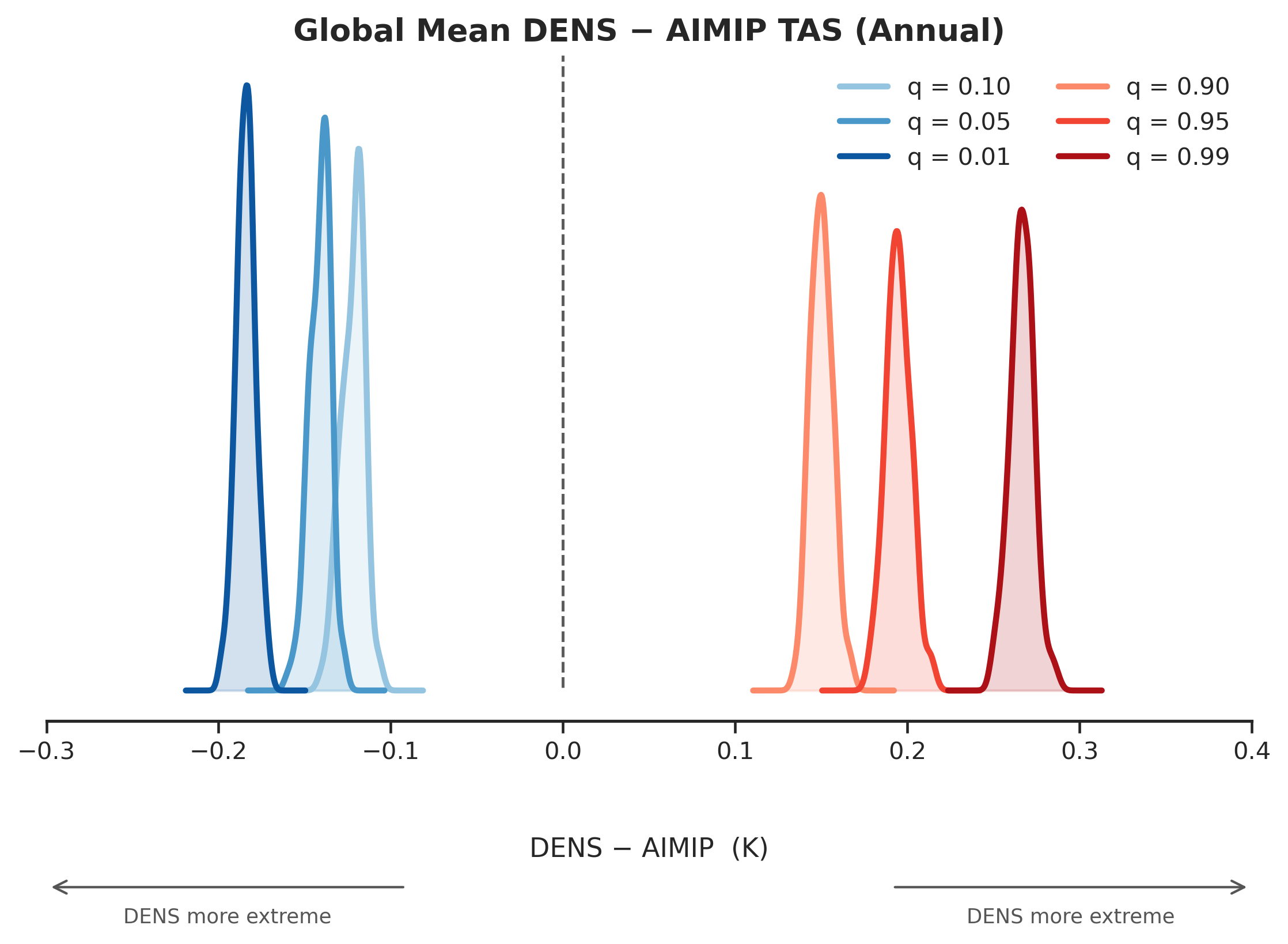}
  \caption[Map and distribution of annual TAS extreme differences]{Same as Figure \ref{fig:extremes}, but for annual TAS.}
  \label{fig:supp_tas_annual_extremes}
\end{figure}

\begin{figure}[ht!]
  \centering
  \includegraphics[width=0.70\textwidth]{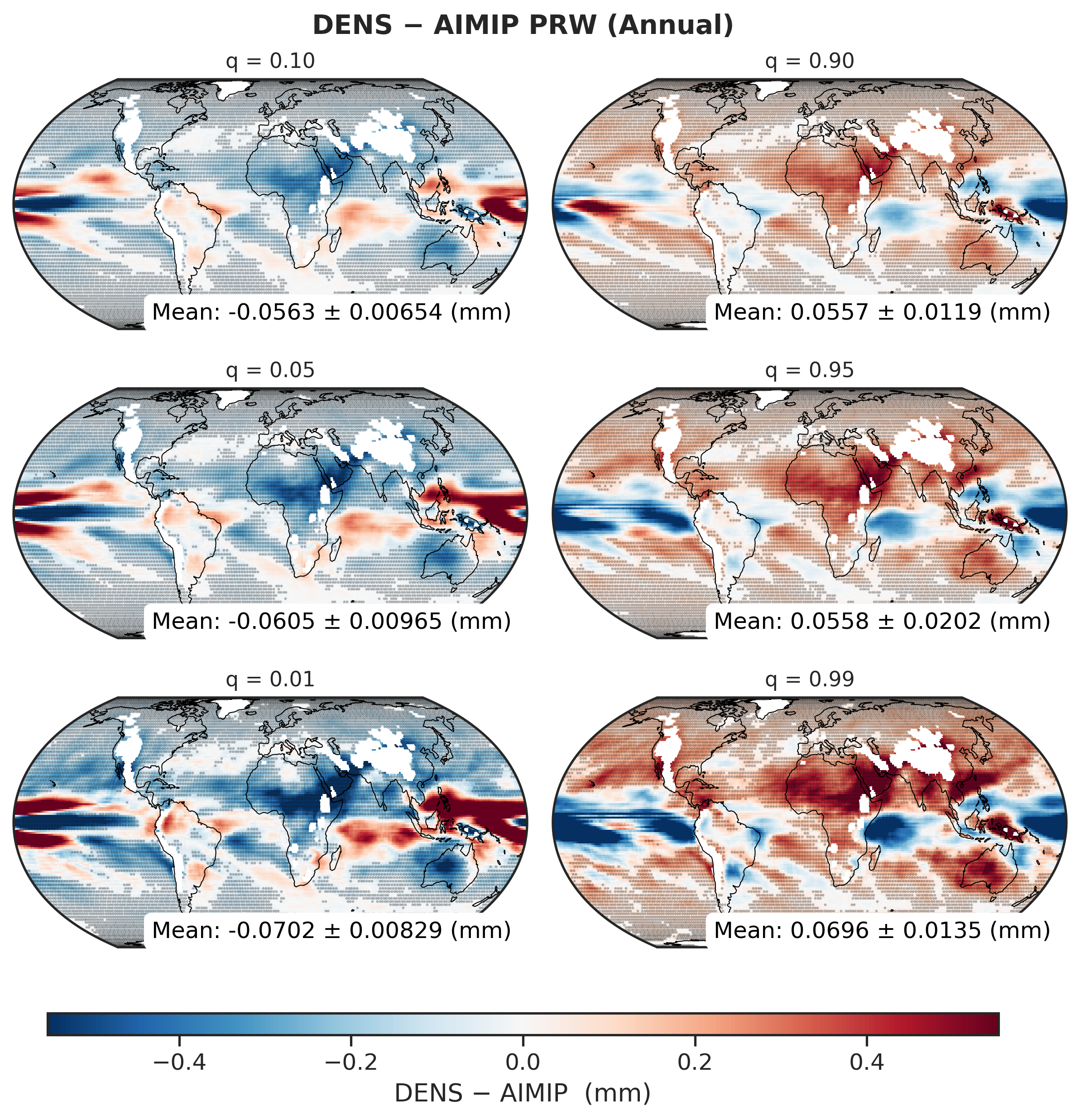}
  \includegraphics[width=0.65\textwidth]{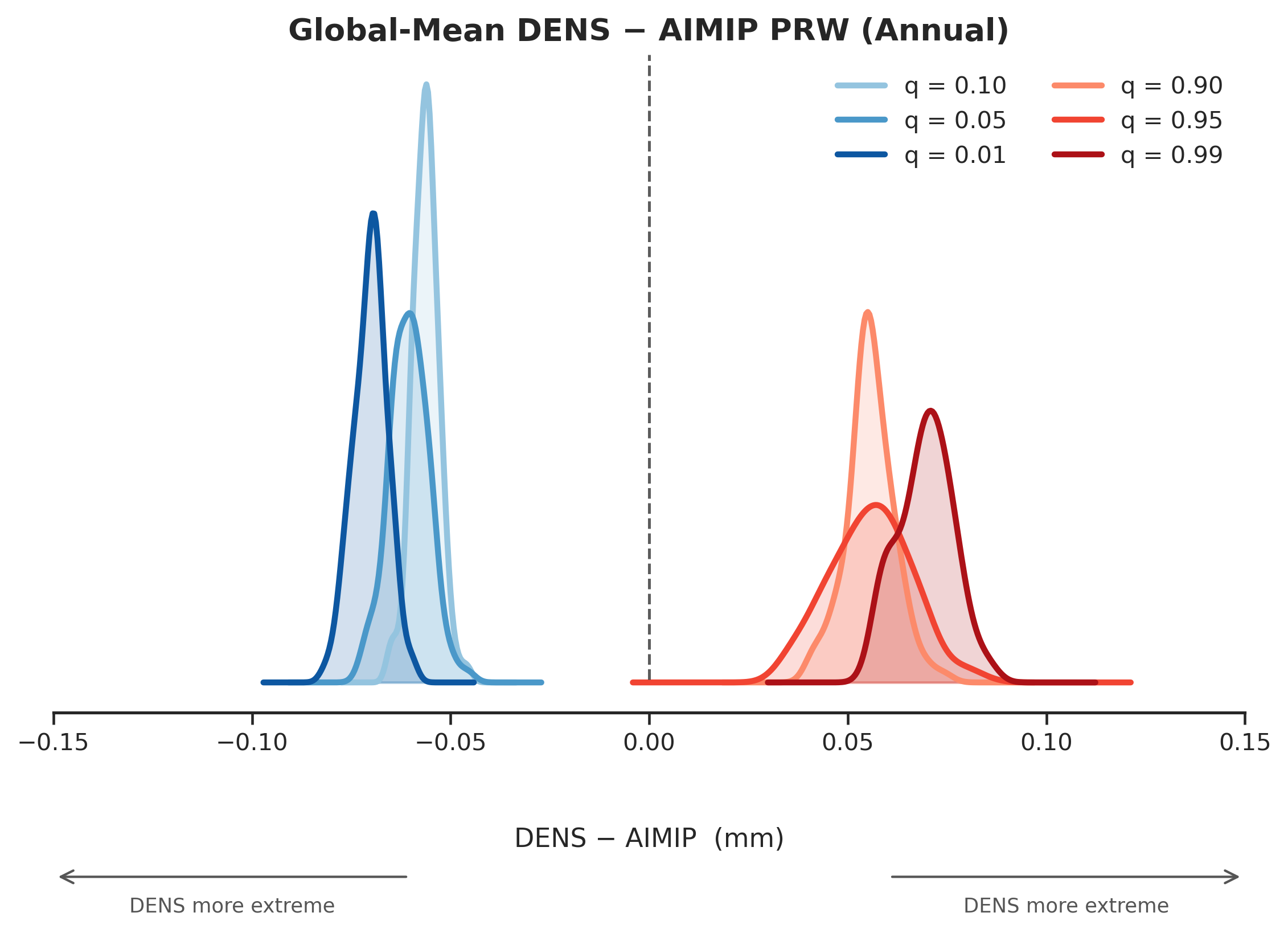}
  \caption[Map and distribution of annual PRW extreme differences]{Same as Figure \ref{fig:extremes}, but for annual PRW.}
  \label{fig:supp_prw_annual_extremes}
\end{figure}

\begin{figure}[ht!]
  \centering
  \includegraphics[width=0.70\textwidth]{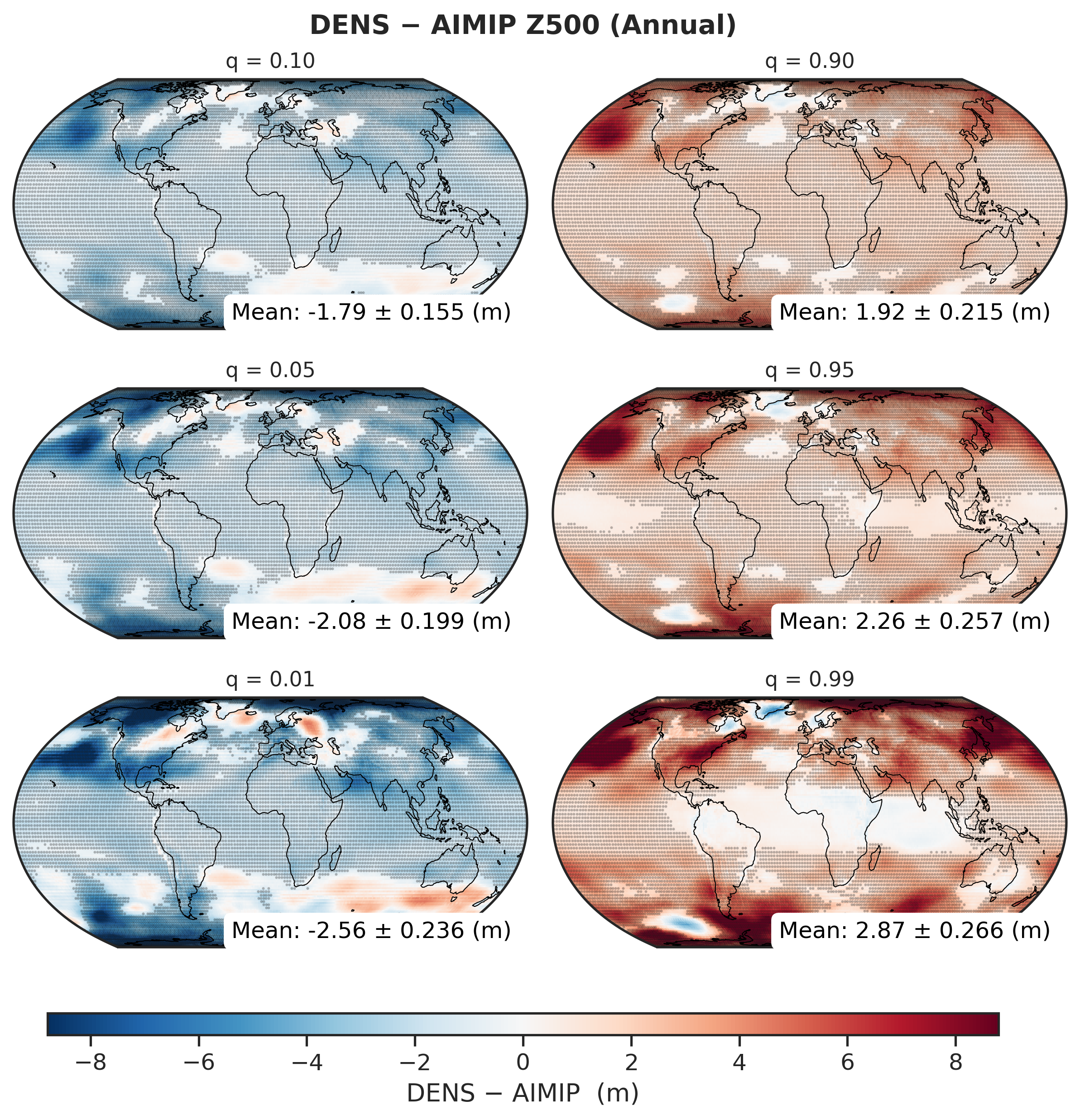}
  \includegraphics[width=0.65\textwidth]{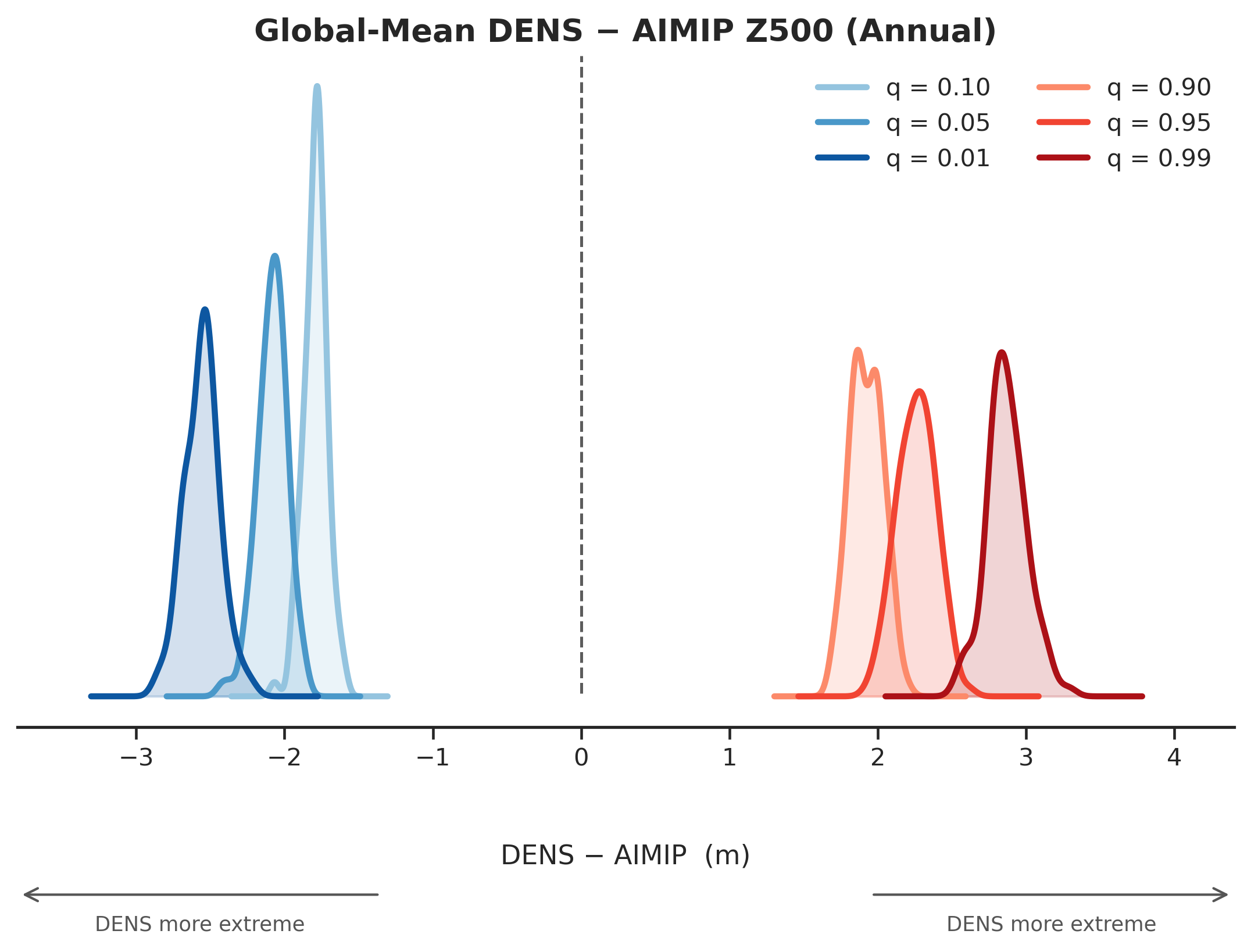}
  \caption[Map and distribution of annual Z500 extreme differences]{Same as Figure \ref{fig:extremes}, but for annual Z500.}
  \label{fig:supp_z500_annual_extremes}
\end{figure}

\clearpage

\section{Supplemental PIT dispersion maps}\label{sec:supp_pit_dispersion}

\begin{figure}[ht!]
  \centering
  \includegraphics[width=0.6\textwidth]{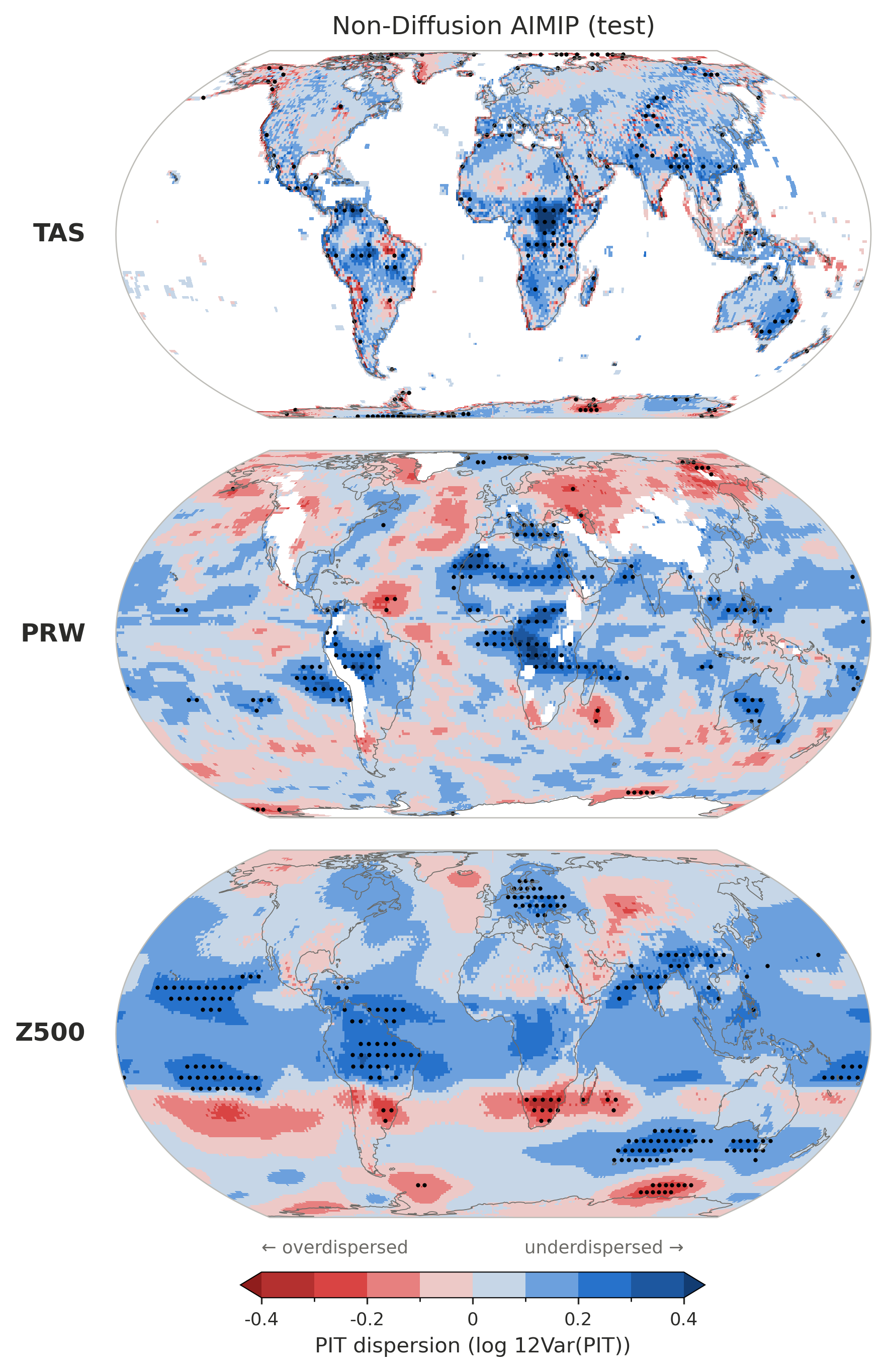}
  \caption[PIT dispersion, AIMIP test period]{As in Figure \ref{fig:dispersion}, but for non-diffusion AIMIP models over the test period (2015--2024).}
  \label{fig:dispersion_test}
\end{figure}

\begin{figure}[ht!]
  \centering
  \includegraphics[width=0.9\textwidth]{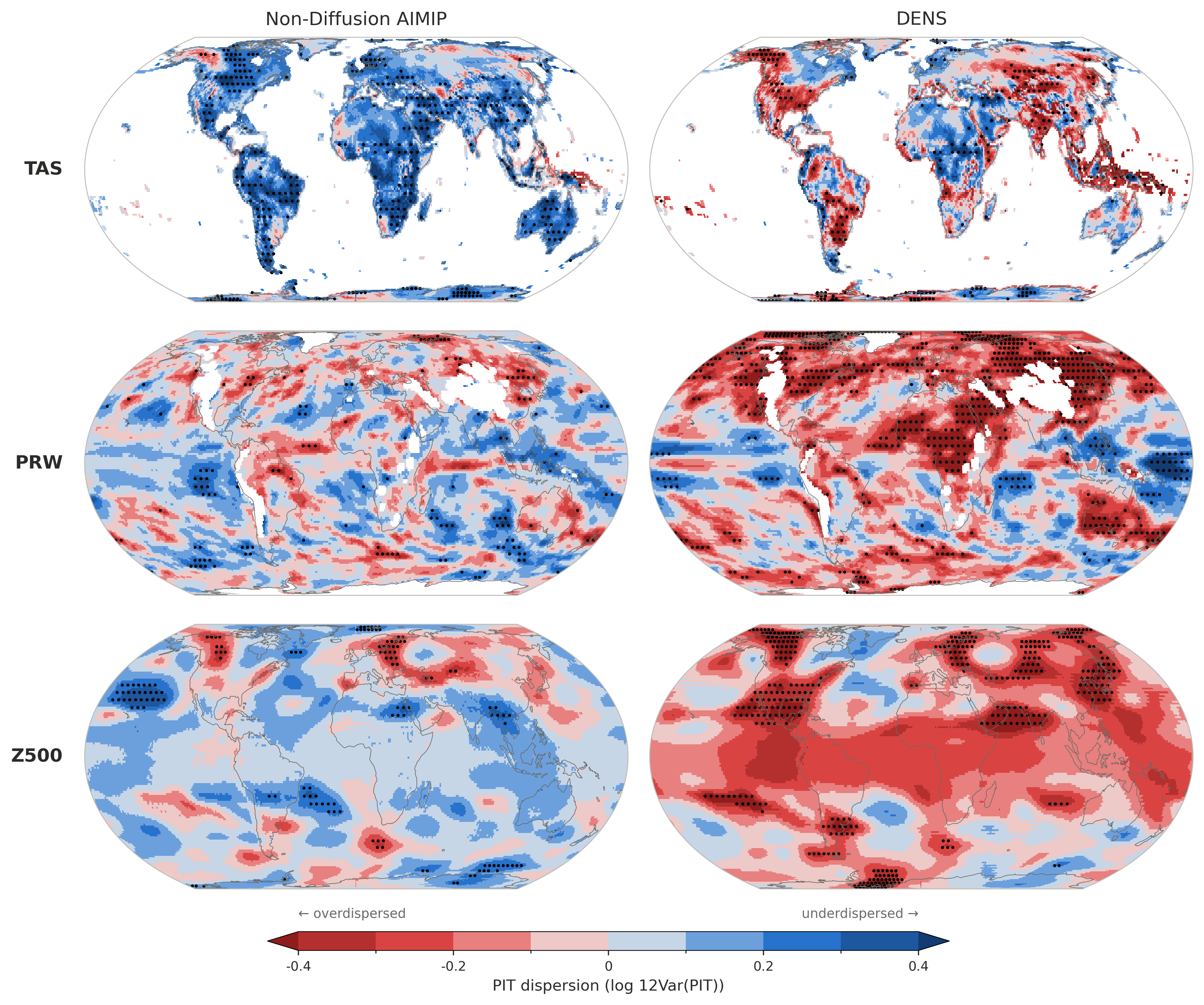}
  \caption[PIT dispersion, 12-month smoothed]{As in Figure \ref{fig:dispersion}, but using 12-month smoothed data over the training period (1979--2014).}
  \label{fig:dispersion_12month}
\end{figure}
\clearpage

\section{PIT D-metric Significance}\label{sec:dmetric_bootstrap}

To measure deviations from marginal calibration, we report the D-metric (see Nipen and Stull~\citep{Nipen} and Equation \ref{eq:caldev}).

\begin{equation}
  D = \sqrt{\frac{1}{B}\sum_{i=1}^{B}
      \left(\frac{b_{i}}{\lVert T \rVert} - \frac{1}{B}\right)^{2}},
  \label{eq:caldev}
\end{equation}
Here, $B$ is the number of bins (in our case 10), $b_{i}$ is the number of ERA5 outputs assigned to bin $i$, and $\lVert T \rVert$ is the total number of ERA5 samples. We apply cosine-latitude weighting when forming $\frac{b_{i}}{\lVert T \rVert}$.

To evaluate whether the calculated D-metric is significant beyond sampling noise, we form a null distribution of D-metrics by selecting a single member to treat as pseudo-observations. We bin this member by deciles formed by its framework with that member excluded, compute D, and repeat for all members. The observed D is bolded if it exceeds the 90th percentile of this null.

\begin{table*}[ht!]
  \centering
  \caption[PIT D-metric bootstrapping results]{Deviation metrics associated with PIT Histograms in Figure \ref{fig:pit}. Bolded values indicate a greater deviation than expected deviations due to sampling noise in 90\% or more of the bootstrap iterations.}
  \label{tab:pitd_bootstrap}
  \includegraphics[width=0.95\textwidth]{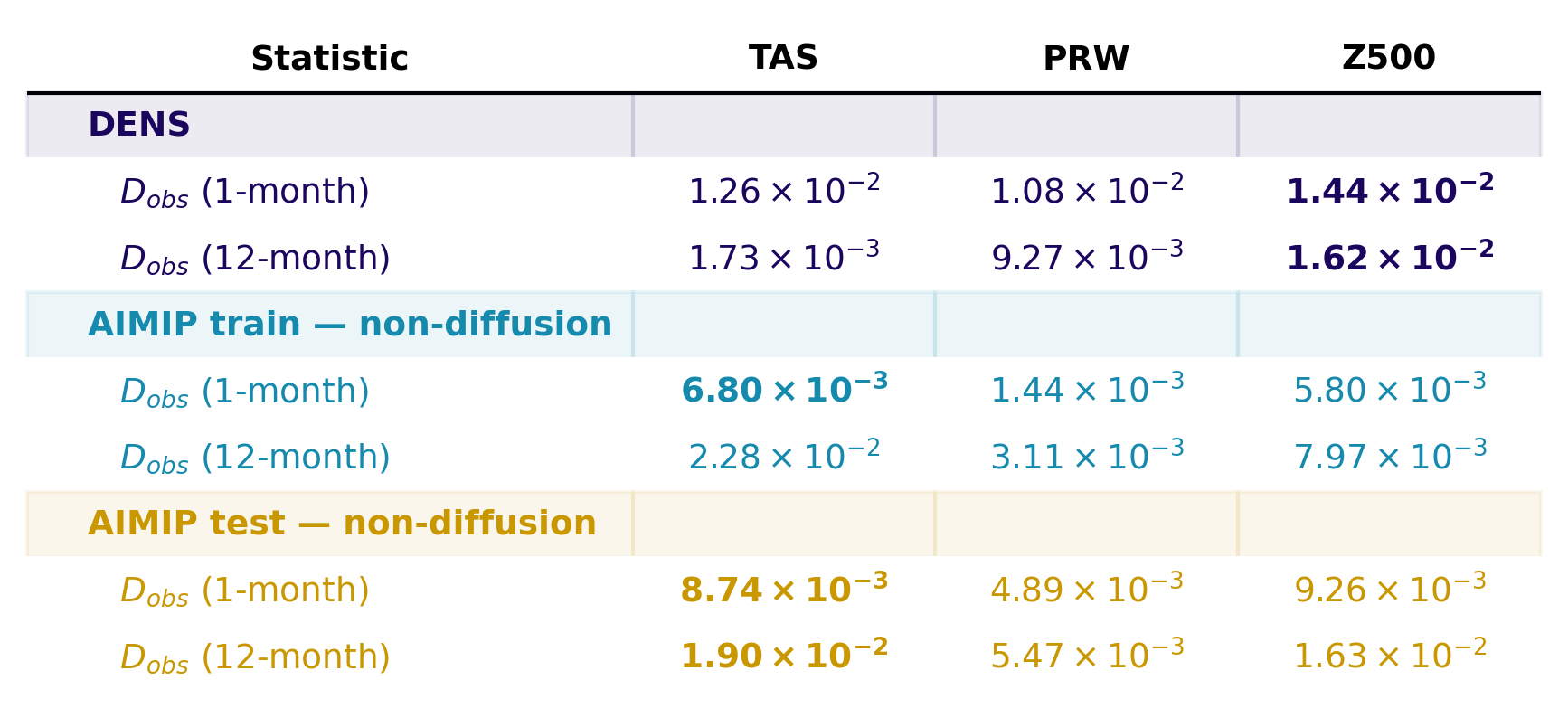}
\end{table*}
\clearpage

\section{Supplemental PIT histograms}\label{sec:diff_pit}

\begin{figure}[ht!]
  \centering
  \includegraphics[width=0.75\textwidth]{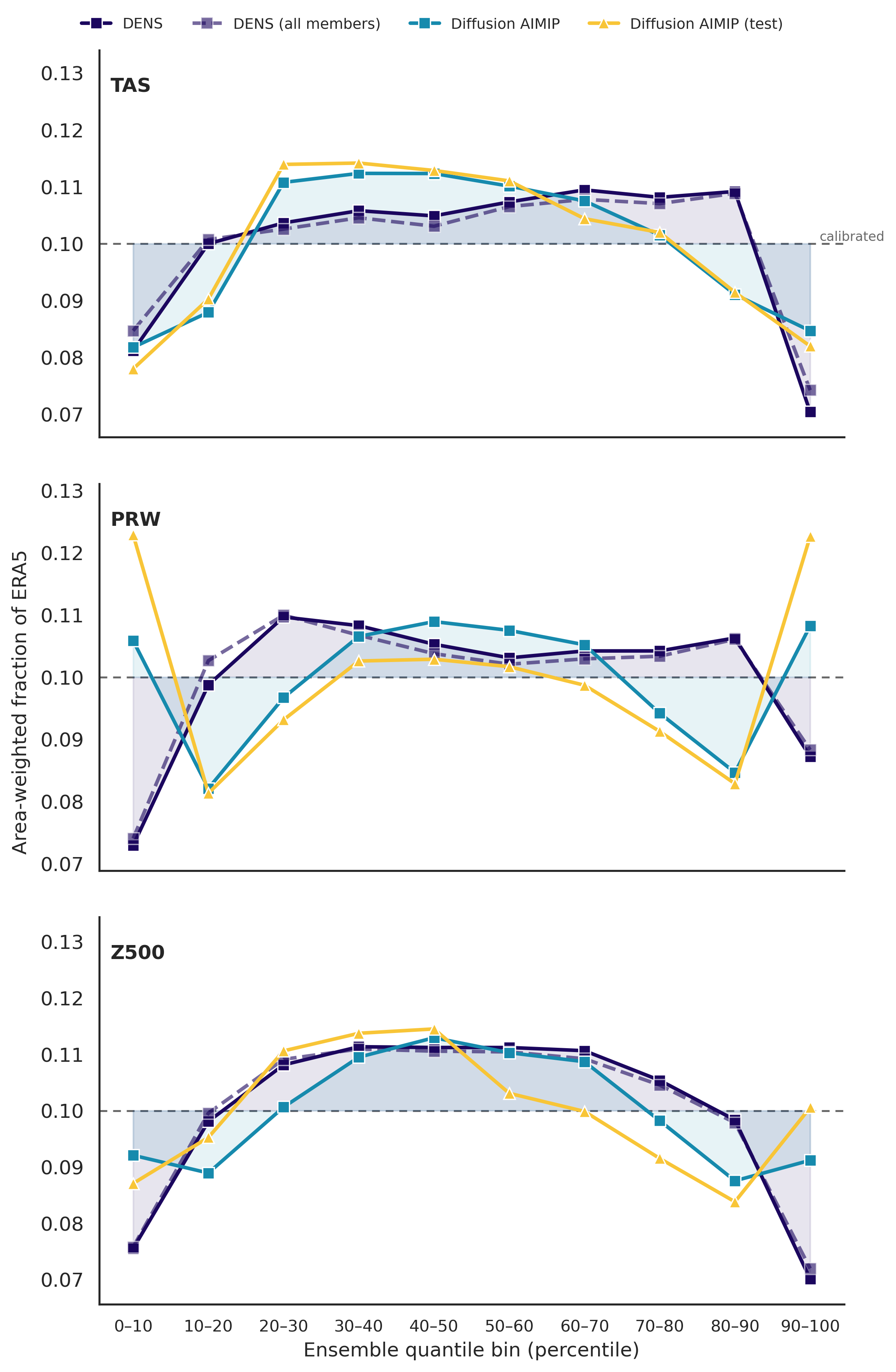}
  \caption[Global PIT for AIMIP diffusion models]{As in Figure \ref{fig:pit}, but for AIMIP diffusion models.}
  \label{fig:diff_pit}
\end{figure}

\begin{figure}[ht!]
  \centering
  \includegraphics[width=0.75\textwidth]{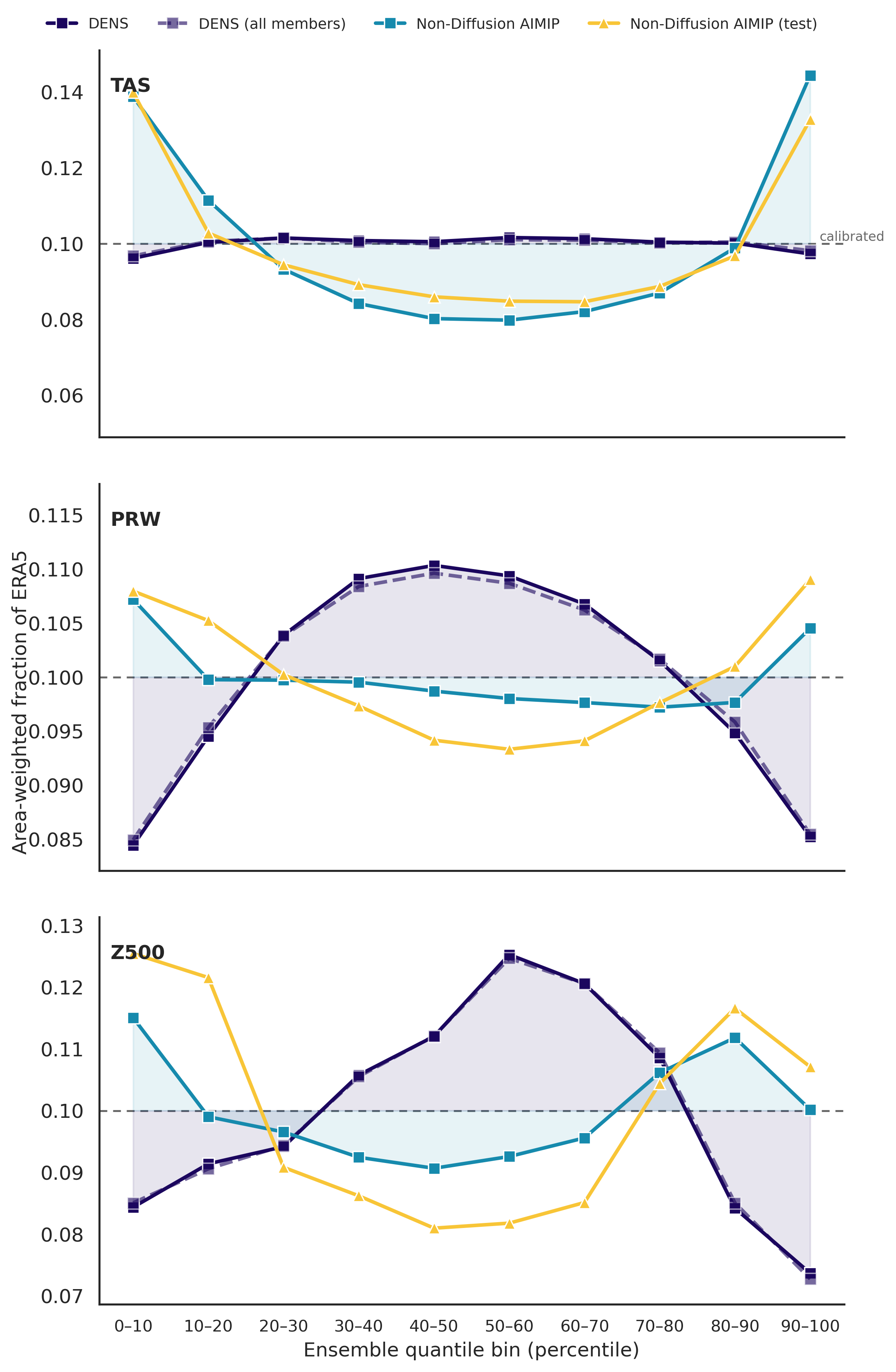}
  \caption[Global 12-month PIT for AIMIP non-diffusion models]{As in Figure \ref{fig:pit}, but using 12-month smoothed data.}
  \label{fig:diff_pit_12_month}
\end{figure}
\clearpage

\section{Variance Decomposition}\label{sec:supp_temporal_var}

Let Y represent a climate variable output for an arbitrary location from a multi-model ensemble. Informally, we use the law of total variance to decompose the total variance of Y across models, members, and time into a sum of model- and time-mean intramodel variance (internal variability), time-mean intermodel variance (model uncertainty), and the temporal variance of the multi-model mean (forced variability): 

\begin{equation}
  \begin{split}
    \text{Var}(Y) & =\mathbb{E}[\text{Var}(Y \mid T)]+\text{Var}(\mathbb{E}[Y \mid T]) \\
& =  \mathbb{E}[\mathbb{E}[\text{Var}(Y \mid M, T)\mid T]] + \mathbb{E}[\text{Var}(\mathbb{E}[Y \mid M, T] \mid T)] + \text{Var}(\mathbb{E}[Y \mid T]) \\
& = \mathbb{E}[\text{Var}(Y \mid M, T)] + \mathbb{E}[\text{Var}(\mathbb{E}[Y \mid M, T] \mid T)] + \text{Var}(\mathbb{E}[Y \mid T]) \\
& = \mathbb{E}[\text{intramodel variance}] + \mathbb{E}[\text{intermodel variance}] + \text{forced variability}
  \end{split} 
\end{equation}

where M represents model and T represents time step. Between lines 2 and 3, we use the law of total expectation to write $\mathbb{E}[\mathbb{E}[\text{Var}(Y \mid M, T)\mid T]]=\mathbb{E}[\text{Var}(Y \mid M, T)]$. 

Since we remove the time-mean of each model, the intermodel and forced variance terms can be combined into a single term, representing the mean of single-model forced variance across models. 

\begin{equation}
  \begin{split}
    \text{Var}(Y) & =\mathbb{E}[\text{Var}(Y \mid M)]+\text{Var}(\mathbb{E}[Y \mid M]) \\
& = \mathbb{E}[\text{Var}(Y \mid M)] \\ 
& =  \mathbb{E}[\mathbb{E}[\text{Var}(Y \mid M, T)\mid M]] + \mathbb{E}[\text{Var}(\mathbb{E}[Y \mid M, T] \mid M)]  \\
& =  \mathbb{E}[\text{Var}(Y \mid M, T)] + \mathbb{E}[\text{Var}(\mathbb{E}[Y \mid M, T] \mid M)]  \\
& = \mathbb{E}[\text{intramodel variance}] + \mathbb{E}[\text{single-model forced variance}]
  \end{split}
\end{equation}
\clearpage

\section{Small-Sample Bias Correction}\label{sec:supp_biascorr}

In this work, we use sample estimates from finite ensembles, where we have $m$ models with $n = 5$ members each ($m = 5$, except for AIMIP PRW, where $m = 4$ as DLESyM is omitted). Denoting the sample estimates of internal variability, model uncertainty, and forced variability by $\hat{I}$, $\hat{M}$, and $\hat{F}$, their expected values under a balanced random-effects model are
\begin{equation}
  \begin{split}
    \mathbb{E}[\hat{I}] &= \sigma^2_I, \\
    \mathbb{E}[\hat{M}] &= \sigma^2_M + \frac{\sigma^2_I}{n}, \\
    \mathbb{E}[\hat{F}] &= \sigma^2_F + \frac{\sigma^2_M}{m} + \frac{\sigma^2_I}{m n},
  \end{split}
\end{equation}
where $\sigma^2_I$, $\sigma^2_M$, and $\sigma^2_F$ are the population internal variability, model uncertainty, and forced variability. On the other hand, $\frac{\sigma^2_I}{n}$ and $\frac{\sigma^2_M}{m} + \frac{\sigma^2_I}{m n}$ are terms from sampling bias with a finite ensemble. 

\begin{figure}[ht!]
  \centering
  \panelgraphic[north east]{0.95\textwidth}{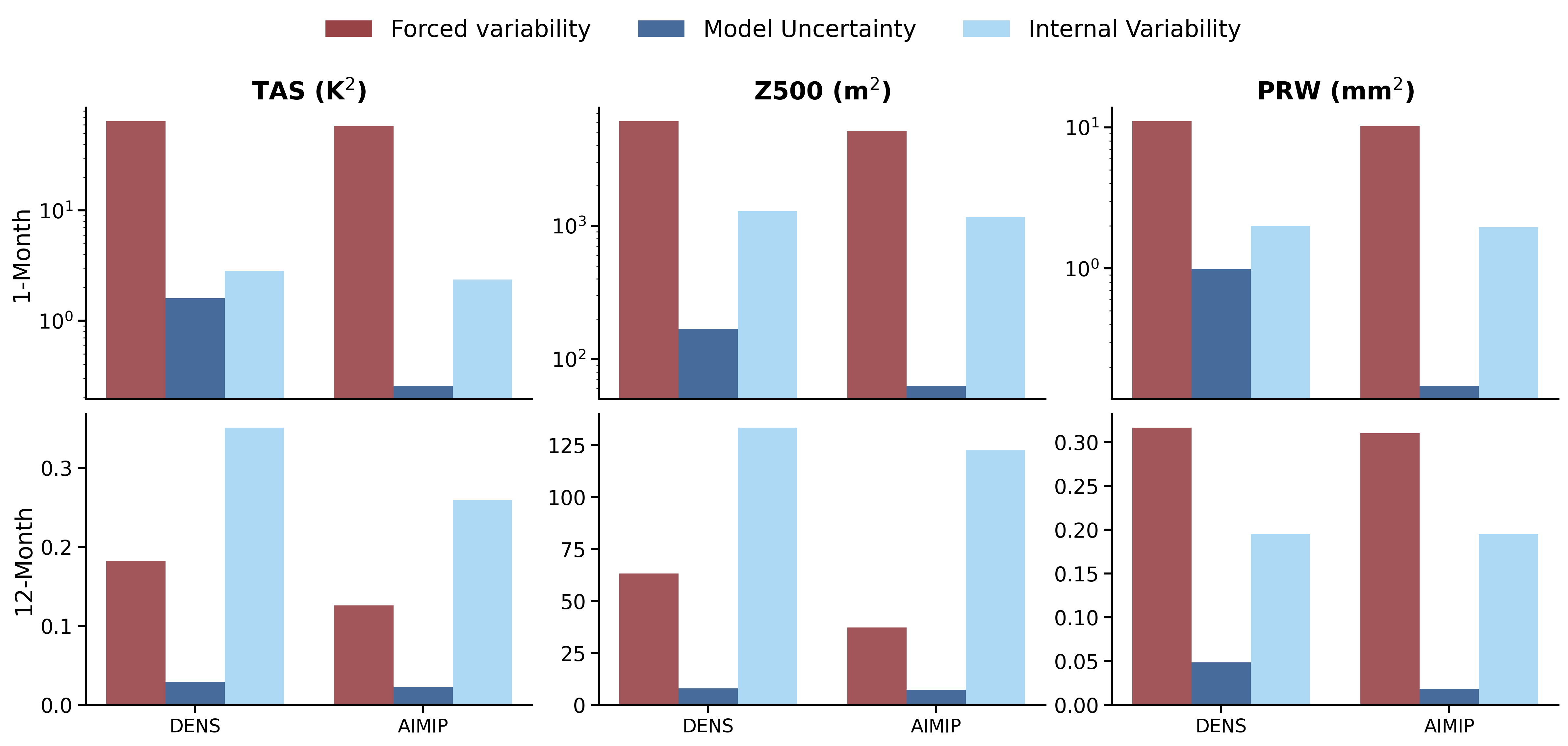}{-0.004/0.862/a, -0.004/0.443/b}
  \caption[Bias-Corrected Variance Components]{Same as Figure \ref{fig:smoothing}, but with model uncertainty and forced variability corrected for small-sample bias (Section \ref{sec:supp_biascorr}). Both (a) monthly and (b) annual (12-month smoothed) results are shown over the training period.}\label{fig:variance_components_biascorr}
\end{figure}

The variance components with sampling error biases subtracted are shown in Figure \ref{fig:variance_components_biascorr}. The correction reduces model uncertainty and slightly reduces forced variability in both frameworks. The qualitative conclusions of the main text are unchanged: DENS exhibits larger internal variability, model uncertainty, and forced variability than AIMIP.

\clearpage

\section{ Quantile Difference Attribution}\label{supp:sec_shapley}

We remark that, since the quantile differences in Figure \ref{fig:extremes} are with respect to the model-mean quantiles of each framework, the ``forced variability'' considered in the Shapley attribution below is related to the mean single-model forced variance. This depends on the variability of the multi-model mean but is also influenced by model uncertainty. That is to say, the Shapley contributions of forced variability below are not directly comparable to the forced variability shown in Figure \ref{fig:smoothing}, as they are also influenced by model uncertainty. Expected single-model forced variance is roughly equivalent to forced variability on the monthly time scale, as forced variability is by and large the dominant term monthly (Figure \ref{fig:smoothing}). However, for the annually smoothed data, model uncertainty and forced variability both contribute to the observed single-model forced variance, particularly for TAS and Z500.

In section \ref{supp:sec_shapley_results} we show global-mean and spatial Shapley values for both forced and internal variability at monthly and annual time scales. We include background on the Shapley value in section \ref{supp:sec_shapley_background} and its relation to factor separation in section \ref{supp:sec_factor_separation}. We detail our implementation of quantile difference attribution in section \ref{supp:sec_shapley_implementation}. 

\subsection{ Shapley Results}\label{supp:sec_shapley_results}

Since we perform attribution by constructing paths from DENS to AIMIP (see Supplemental Section \ref{supp:sec_shapley_implementation}), the resultant Shapley values are related to the AIMIP minus DENS quantile difference, which is the opposite sign of Figure \ref{fig:extremes}. Hence, a negative Shapley value indicates that the component — either forced or internal variability — contributes to a positive quantile difference in Figure \ref{fig:extremes}; while a positive Shapley value denotes a negative contribution.

\begin{figure}[ht!]
  \centering
  \includegraphics[width=\textwidth]{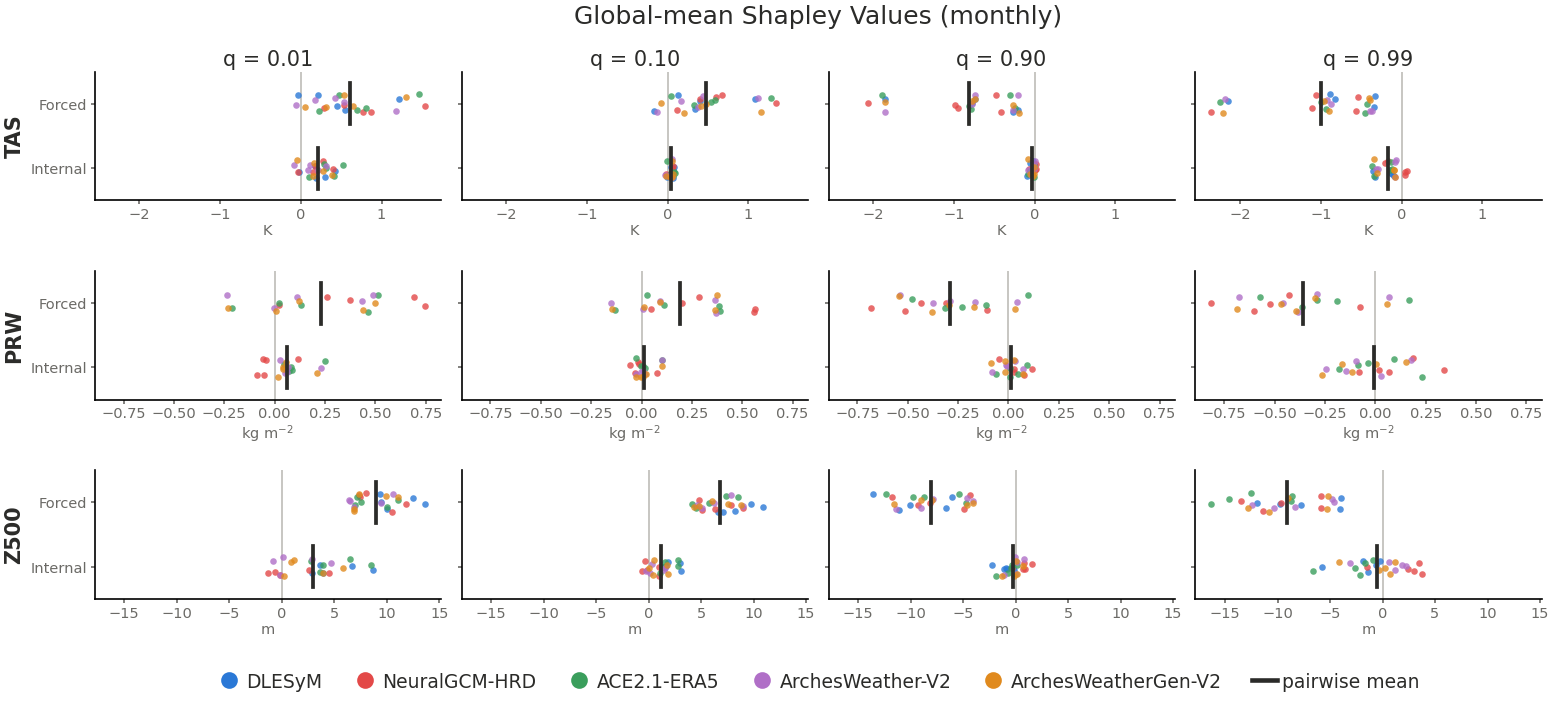}

  \caption[Global-Mean Shapley Values, Monthly]{Global-mean pairwise Shapley values of monthly forced and internal variability, with respect to AIMIP-DENS quantile differences (over the training period), along with pairwise mean Shapley values.}
  \label{fig:shapley_global_monthly}
\end{figure}

\begin{figure}[ht!]
  \centering
  \panelgraphic{0.85\textwidth}{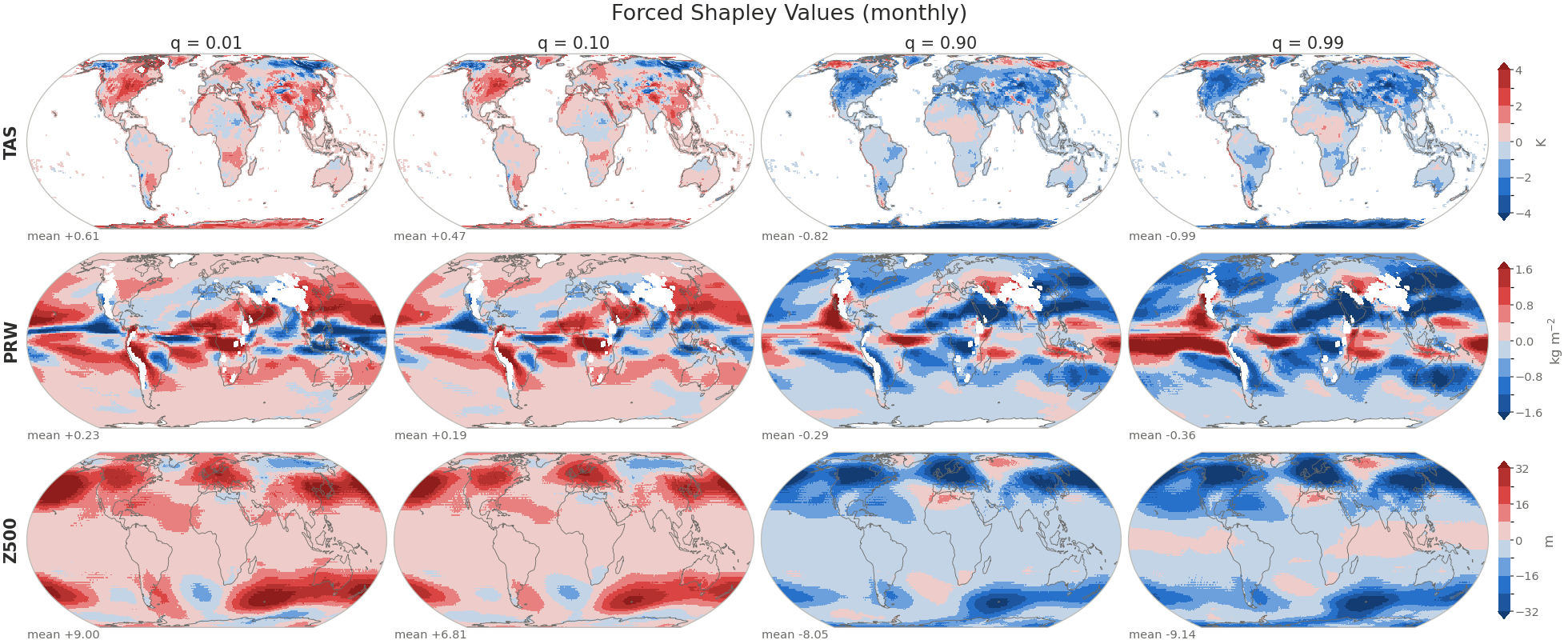}{0/1/a}

  \vspace{0.5em}
  \panelgraphic{0.85\textwidth}{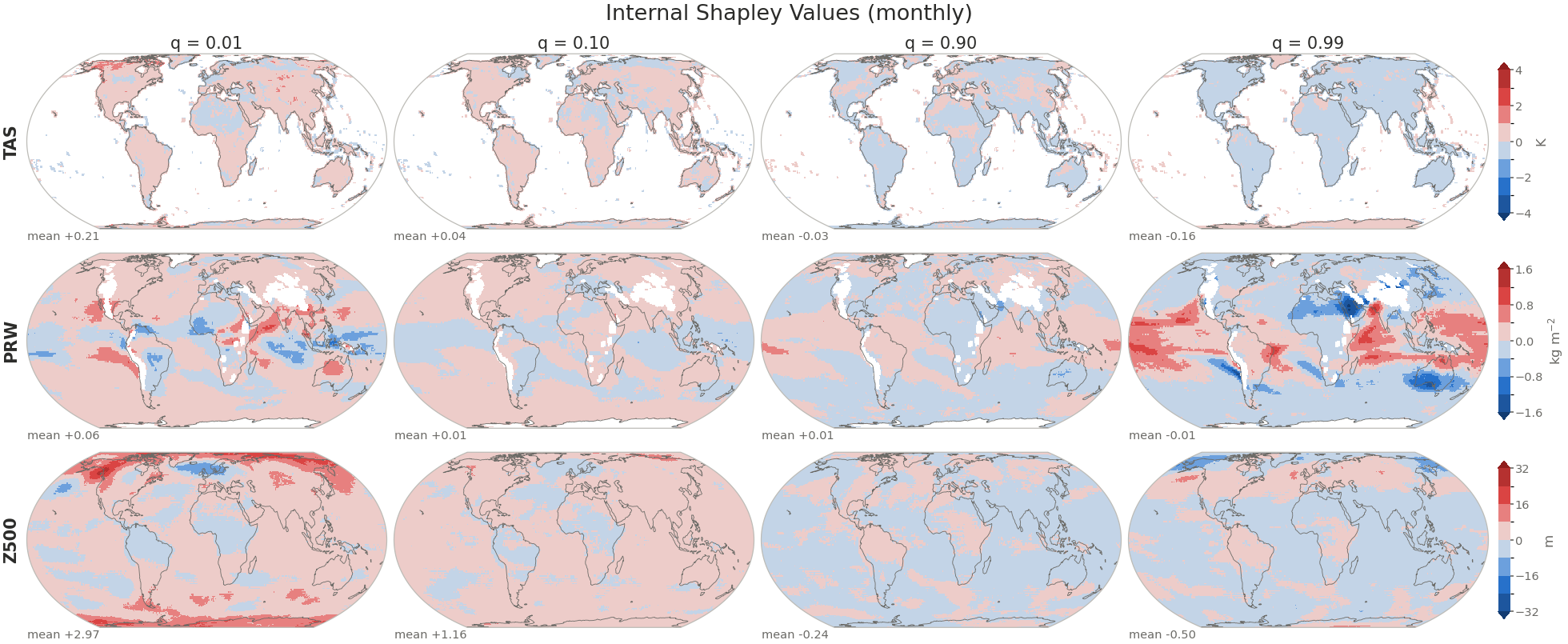}{0/1/b}

  \caption[Shapley Value Maps, Monthly]{Pairwise-mean Shapley values of monthly (a) forced and (b) internal variability.}
  \label{fig:shapley_global_monthly_maps}
\end{figure}
\clearpage

\begin{figure}[ht!]
  \centering
  \includegraphics[width=\textwidth]{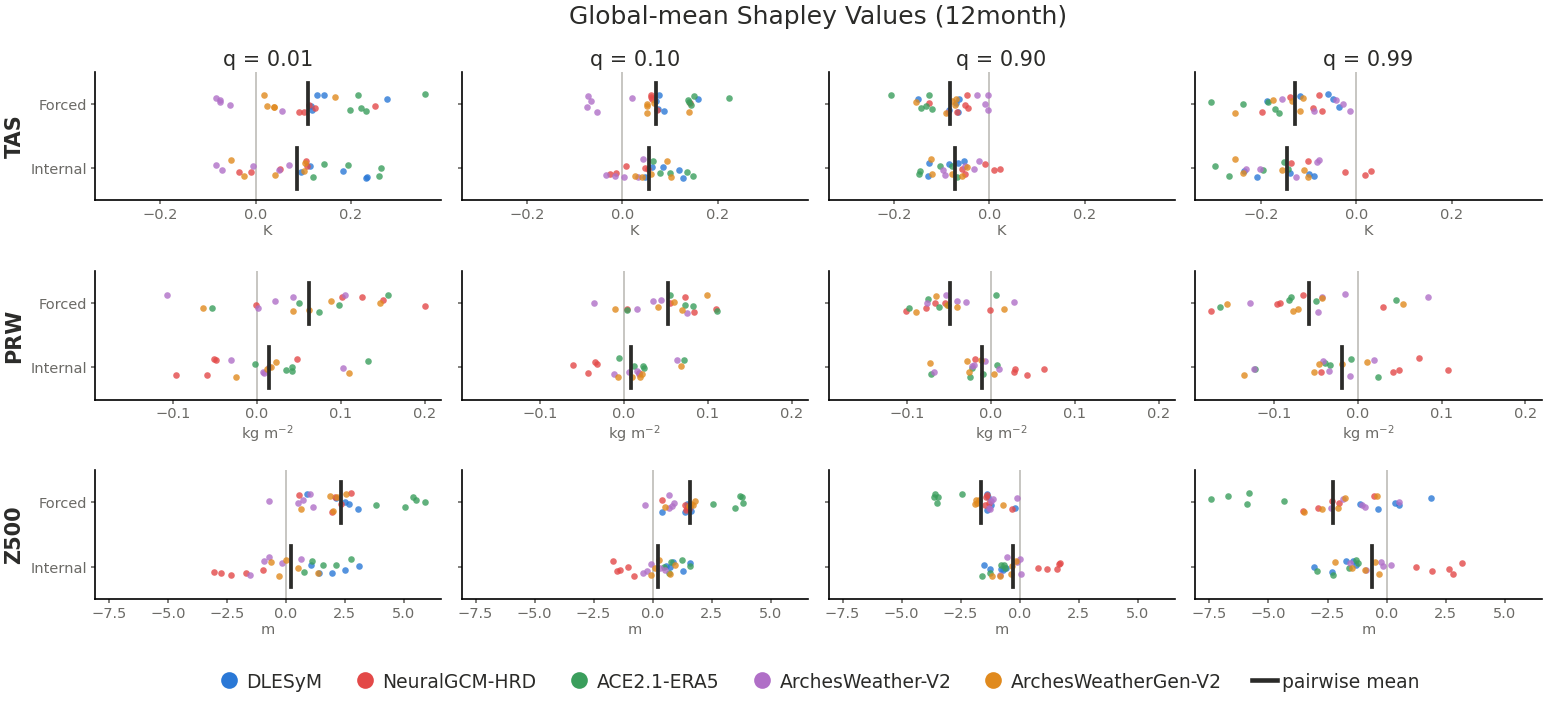}

  \caption{As in Figure \ref{fig:shapley_global_monthly} but for annually smoothed data.}
  \label{fig:shapley_global_annually}
\end{figure}

\begin{figure}[ht!]
  \centering
  \panelgraphic{0.85\textwidth}{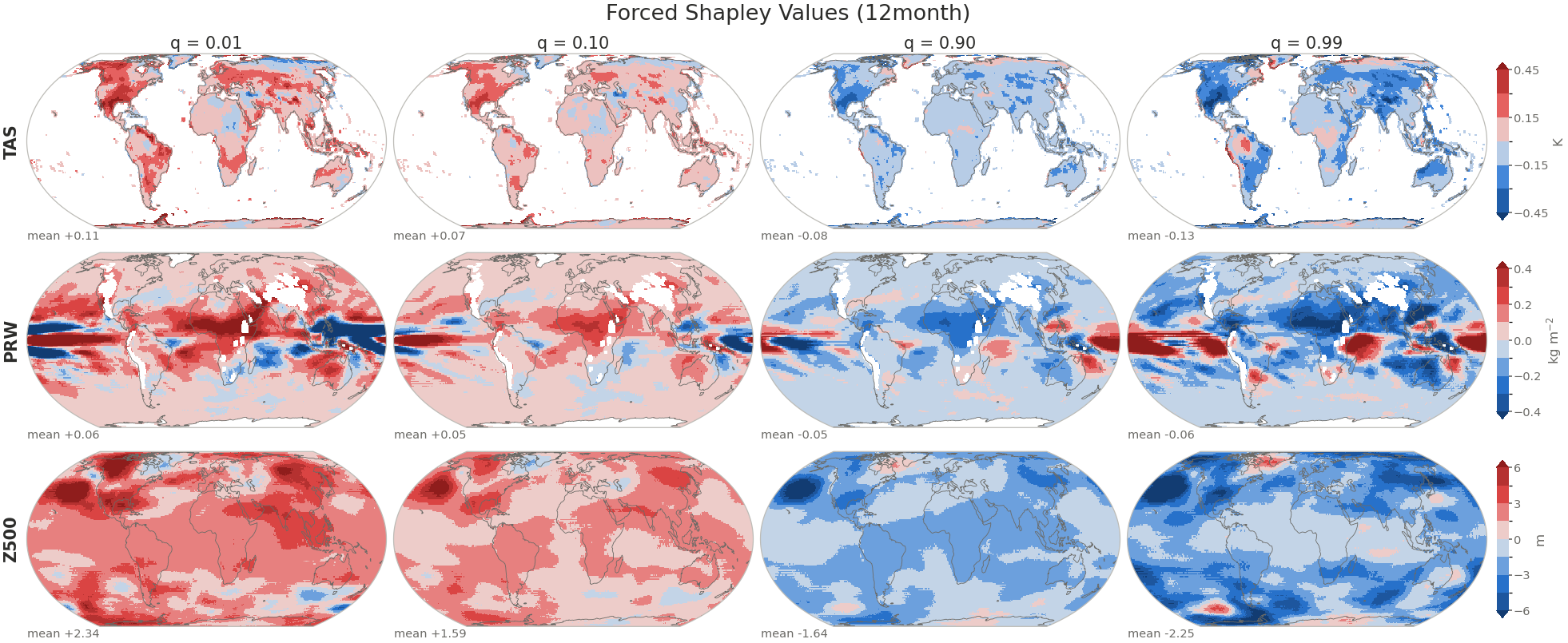}{0/1/a}

  \vspace{0.5em}
  \panelgraphic{0.85\textwidth}{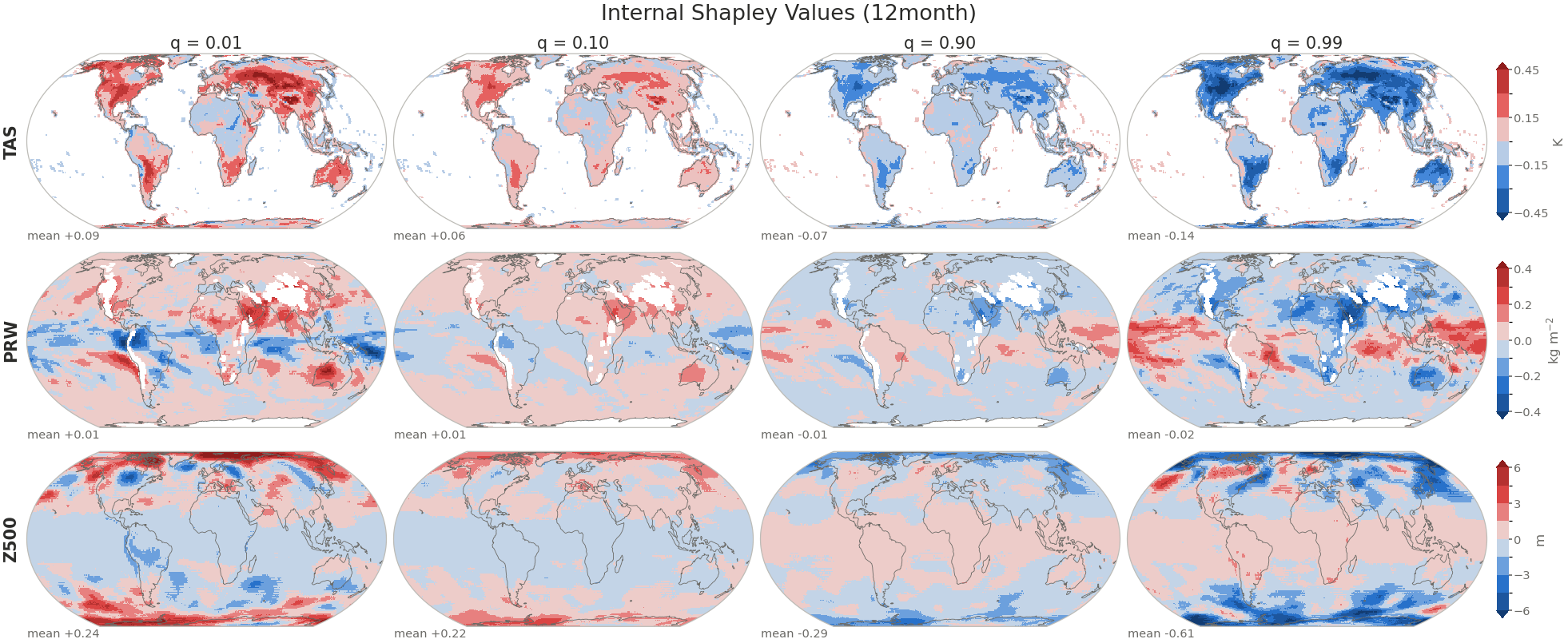}{0/1/b}

  \caption{As in Figure \ref{fig:shapley_global_monthly_maps} but for annually smoothed data.}
  \label{fig:shapley_global_annually_maps}
\end{figure}
\clearpage

\subsection{ Shapley Value}\label{supp:sec_shapley_background}

In Figure \ref{fig:extremes} we find that AIMIP produces weaker extremes than DENS. Here, we seek to attribute the observed quantile discrepancy to differences in internal and forced variability between the two frameworks. For this, we make use of the Shapley value from cooperative game theory \citep{Shapley1953}. A cooperative game consists of a set of players $\mathcal{N}$, called a grand coalition, and a value function $v(S)$ which assigns payoffs to coalitions of players $S \subseteq \mathcal{N}$. A player $i$'s marginal contribution to a coalition $S$ is the value gained by adding player $i$ to the coalition, $v(S \cup \{i\}) - v(S)$. 

The Shapley value is a solution concept, or a rule determining how the payoff of the grand coalition $v(\mathcal{N})$ is divided up among the players. The Shapley value $\phi_i$ considers all possible permutations of players, and rewards player $i$ based on their average marginal contribution to the coalition preceding $i$ in each permutation:  

\begin{equation}
    \phi_i(v) = \sum_{S \subseteq \mathcal{N} \backslash \{i\}} \frac{|S|! \ (|\mathcal{N}| - |S| - 1)!}{|\mathcal{N}|!} \ (v(S \cup \{i\}) - v(S))
\end{equation}

That is to say, to compute a player's Shapley value, we consider all coalitions which do not include the player and calculate the player’s marginal contribution to each. We then average these contributions, weighting each coalition by the frequency at which it precedes the player across all permutations, $\frac{|S|! \ (|\mathcal{N}| - |S| - 1)!}{|\mathcal{N}|!}$. 

The Shapley value is the only solution concept satisfying the four axioms of fairness in cooperative game theory: efficiency, symmetry, null player, and linearity \citep[see][for more details]{Shapley1953}. The Shapley value has been used extensively in explainable AI for fair feature attribution \citep{shapley_value}, including various applications to weather and climate \citep[e.g.,][]{Yang2024}. 

\subsection{ Relation to Factor Separation}\label{supp:sec_factor_separation}

The Shapley value is closely related to factor separation in numerical climate modeling \citep{shapley_factor_separation}. Factor separation is a family of methods for attributing changes in climate model output, relative to a baseline state, to changes in model components \citep{Lunt2021}. These components take the form of binary factors, representing, e.g., whether prescribed CO2 is from a preindustrial baseline or a warming scenario. 

Factor separation involves running simulations for subsets of factors. Attribution is performed by considering non-returning paths from the baseline state to the perturbed state, incorporating one additional factor at a time. For a given path, each factor's contribution is measured as the change in model output relative to the preceding subset of factors along the path. For $N$ factors, the linear-sum method of factor separation introduced by Lunt et al.~\citep{Lunt2021} computes contributions with respect to all $N!$ paths. The attribution of each factor is defined as its average contribution over all paths \citep{Lunt2021}, which is equivalent to the factor's Shapley value \citep{shapley_factor_separation}; here, the value function is given by the change in model output relative to baseline. The linear-sum method is an extension of the technique outlined in Lunt et al.~\citep{Lunt2012}, which also averages factor contributions over all subsets of remaining factors, but weights each subset equally. This is equivalent to the Banzhaf value, which equals the Shapley value for $N=2$ factors but does not satisfy the efficiency axiom generally \citep{banzhaf, Shapley1953}. The lack of efficiency for $N>2$ was the motivation in Lunt et al.~\citep{Lunt2021} for extending the technique, yielding a revised method equivalent to the Shapley value. 

Our implementation is distinct from traditional factor separation methods, as we do not run new AIMIP or DENS simulations for each subset of factors. Instead, as outlined below, we exchange components of variability between the two ensembles post hoc, creating synthetic hybrid ensembles to construct paths from DENS to AIMIP. 

\subsection{ Implementation}\label{supp:sec_shapley_implementation}

For the $q$th quantile at a given grid point, we rewrite the model-mean quantile difference $\Delta Q$ as an average of pairwise differences across all AIMIP-DENS model pairs. Note that this is the negative of the quantile difference shown in Figure \ref{fig:extremes}.

\begin{equation}
    \Delta Q = \frac{1}{N_A N_D}\sum_{i=1}^{N_A}\sum_{j=1}^{N_D} \left(Q_A^{(i)} - Q_D^{(j)}\right)
    \label{eqn:pairwise_quantile_diff}
\end{equation}
where $N_A$ denotes the number of AIMIP models (5 for TAS and Z500, 4 for PRW), $N_D$ denotes the number of DENS models (5 for each variable), and $Q_A^{(i)}$ and $Q_D^{(j)}$ denote the $q$th quantile of the $i$th AIMIP and $j$th DENS model respectively.

We write $Q(F, I)$ to mean the $q$th quantile of the time series obtained by adding internal variability $I$ to forced variability $F$. Specifically, $F$ represents the ensemble mean of a given model -- belonging to either AIMIP or DENS -- while $I$ is the member-residuals of either the same model or a different model. If $F$ belongs to the same model as $I$, then $Q(F, I)$ is simply that model's $q$th quantile. Otherwise, $Q(F, I)$ is the $q$th quantile of a hybrid ensemble combining one model's member-mean with the other's member-residuals. 

We consider a game with two players $\mathcal{N_A} = \{F_A, I_A\}$, where $F_A = \{F_A^{(i)}\}_{i=1}^{N_A}$ and $I_A = \{I_A^{(i)}\}_{i=1}^{N_A}$ represent the forced and internal variability of each AIMIP model. For each pairwise comparison in equation \ref{eqn:pairwise_quantile_diff}, we define a game $(\mathcal{N_A}, v_{ij})$, where $v_{ij}$ is given by 
\begin{gather}
v_{ij}(S) =
Q(F,I) - Q(F_D^{(j)}, I_D^{(j)}) \label{eqn:pairwise_value} \\
\text{where }
\begin{cases}
F = F_A^{(i)} \text{ if } F_A \in S \text{ else } F_D^{(j)}\\
I = I_A^{(i)} \text{ if } I_A \in S \text{ else } I_D^{(j)}
\end{cases} \notag
\end{gather}

The value function $v_{ij}$ measures the quantile difference between the hybrid ensemble formed by the coalition $S$ and the DENS model $j$. For each pairwise game, we compute the Shapley values of $F_A$ and $I_A$.

\begin{equation}
\phi_{F_A}(v_{ij}) = \frac{1}{2} \left(v_{ij}(\{F_A\}) + v_{ij}(\{F_A, I_A\}) - v_{ij}(\{I_A\})\right) 
\label{eqn:forced_pairwise_shapley}
\end{equation}

\begin{equation}
\phi_{I_A}(v_{ij}) = \left(Q_A^{(i)} - Q_D^{(j)}\right) - \phi_{F_A}(v_{ij}) 
\label{eqn:internal_pairwise_shapley}
\end{equation}

Equation \ref{eqn:internal_pairwise_shapley} follows from the efficiency property of the Shapley value \citep{Shapley1953}, which requires that the Shapley values of $F_A$ and $I_A$ for a model pair sum to the quantile difference between the two models. 

Combining equations \ref{eqn:pairwise_quantile_diff} and \ref{eqn:pairwise_value}, we define a value function for the model-mean quantile difference between the hybrid multimodel ensemble formed by the coalition $S$ and DENS. 

\begin{equation}
    v(S) = \frac{1}{N_A N_D}\sum_{i=1}^{N_A}\sum_{j=1}^{N_D} v_{ij}(S)
    \label{eqn:value}
\end{equation}

By the linearity property of the Shapley value \citep{Shapley1953}, the Shapley values of $F_A$ and $I_A$, with respect to the AIMIP-DENS quantile difference $\Delta Q$, are given by the average of the pairwise Shapley values. 

\begin{equation}
    \phi_{F_A}(v) = \frac{1}{N_A N_D}\sum_{i=1}^{N_A}\sum_{j=1}^{N_D} \phi_{F_A}(v_{ij})
    \label{eqn:forced_shapley}
\end{equation}

\begin{equation}
    \phi_{I_A}(v) = \frac{1}{N_A N_D}\sum_{i=1}^{N_A}\sum_{j=1}^{N_D} \phi_{I_A}(v_{ij})
    \label{eqn:internal_shapley}
\end{equation}

\end{document}